\documentclass[aps,preprint,onecolumn,secnumarabic,nobalancelastpage,amsmath,amssymb,
nofootinbib]{revtex4}

\usepackage[hang,nooneline,scriptsize]{subfigure}
\usepackage{graphics}      % standard graphics specifications
\usepackage{graphicx}      % alternative graphics specifications
\usepackage{longtable}     % helps with long table options
\usepackage{pst-plot}
\usepackage{dcolumn}% Align table columns on decimal point
\usepackage{hyperref}% add hypertext capabilities
\usepackage{bm}            % special 'bold-math' package
\usepackage{mathrsfs}
\usepackage{xcolor}

\begin{document}

\preprint{000/000-000}

\title{Particle Dynamics Around the Black String
}

\author{Sara Rezvanjou${}^{a}$}
\author{Reza Saffari${}^a$}
\email{rsk@guilan.ac.ir}
\author{Mozhgan Masoudi${}^a$}
\author{Saheb Soroushfar${}^{b}$}
%\email{soroush@yu.ac.ir}
\affiliation{${}^a$ Department of Physics, University of Guilan, 41335-1914, Rasht, Iran}
\affiliation{${}^b$ Faculty of Technology and Mining, Yasouj University, Choram 75761-59836, Iran}

\date{\today}

\begin{abstract}
In this paper, some dynamical properties of neutral and charged
particles around a weakly magnetized five-dimensional static black
string have been studied. The perturbation method was also used to
calculate the Innermost Stable Circular Orbit (ISCO) of this metric
in the presence of a magnetic field. The escape velocity of neutral
and charged particles around the black string was derived. In the
next step, the analytical solutions of the equations of motion were
discussed and some possible orbits for particles in the black string
space-time were plotted. Interestingly, it was found that adding an
extra dimension has a slight influence on the effective potential
and one term of the effective force. The magnitude of the new
constant of motion $(J)$ affects both the shape of the potential and
existence of the stable circular orbits. In conclusion, by comparing
a black hole and a black string, it is realized that the value of a
new constant of motion causes slight but interesting differences.
\end{abstract}

\maketitle
 \section{INTRODUCTION}
%%%%%%%%%%%%%%%%%%%%%%%%%%%%%%
Within the last few decades, higher-dimensional space-times have
received special attention in theoretical physics
\cite{ArkaniHamed:1998rs}-\cite{Kaya:2007kh}. Notable attempts to
unify fundamental forces, such as Kaluza and Klein theory, and some
other interesting theories of physics such as String theory or Brane
cosmology, were necessarily formulated in such higher-dimensional
space-times \cite{Lee:2006jx}-\cite{Obers:2008pj}. To investigate
the properties of extra dimensions, studying the higher-dimensional
solutions of Einstein equations (like black strings) is considered
\cite{Lee:2006jx}, \cite{Kunz:2014zxa},
\cite{Kleihaus:2006ee}.\footnote{There are also different methods to
construct black string solutions in 4-dimensional space-time
\cite{Kaloper:1993zc}. For instance, some black string solution in
4-dimensional space-time came from foliation of AdS/dS Schwarzschild
black holes \cite{Emparan:1999fd}, which their stability have been
investigated in \cite{Hirayama:2001bi}.} While there are different
methods to find the exact solution of Einstein gravity in
five-dimensional space-time, the common methods in more than
five-dimensional space-times, are numerical or perturbational
\cite{Kleihaus:2016kxj}. There are also several studies about the
various features of the higher-dimensional space-times, such as the
equation of motion in 5-dimensional space-time
\cite{Mashhoon:1994iu}-\cite{Seahra:2003eb}, the possibility of
detecting extra dimension and its relation to Einstein's Equivalence
Principle \cite{Wesson:1997dq}-\cite{Wesson:2013nk}, the physical
properties of 5-dimensional space-time \cite{Wesson:1994pj}, the
properties of higher-dimensional black holes (charged and rotating
ones) \cite{Kleihaus:2007kc}, \cite{Kunz:2006zz}, the solar system
test of 5-dimensional gravity \cite{Kalligas:1994vf},
\cite{Liu:2000zq}, and even some suggested experimental way to
explore extra dimension by spectroscopy \cite{Luo:2006ad}.

Tangherlini investigated the higher-dimensional black hole for the
first time in 1963 \cite{Tangherlini:1963}. This idea was followed
by Myers-Perry \cite{Myers:1986} for the rotating black hole, and
the simple form of 5-dimensional black string was investigated by
Gregory and Laflamme in 1987 \cite{Gregory:1987nb}, which has been
obtained by adding one non-compact extra dimension to Schwarzschild
black hole and called Black String \cite{Lee:2006jx}. By considering
the non-compact extra dimension the topology of the horizon will be
$ R^{1}\times S^{2}$. There is also another option, adding a compact
flat extra dimension to a Schwarzschild black hole that changes the
horizon topology to $S^{1} \times S^{2}$ and is called a black ring
\cite{Kunz:2014zxa}, \cite{Kleihaus:2016kxj}. With a compact extra
dimension, there is a classification of the spherically symmetric
string-like vacuum solution in 5-dimensional space-time
\cite{Chodos:1980df}. One interesting form of this classification is
a stationary string-like solution of the Einstein equation in (4+1)
dimensions, which has been investigated \cite{Kim:2007ek} and has
shown some interesting features that may help solve the stability
problem(\cite{Marolf:2005vn}, \cite{Horowitz:2001cz}) of the black
string solution. Also, the motion of massive test particles in
static spherically symmetric magnetic-free 5-dimensional space-time
has been investigated \cite{Lacquaniti:2009wc}. Other theories like
Space-Time-Matter (STM) theory \cite{Wesson:1999}-\cite{Liu:2007wx}
consider a non-compact fifth dimension. Some interesting
investigations conducted on the astrophysical implication in the STM
framework \cite{Liu:2000zq}, \cite{Liu:2007wx}, \cite{Liu:1996hs},
\cite{Liu:1992bp}. There is also another kind of cosmological model
in 5 dimensions called brane world (Membrane theory)
\cite{Brax:2003fv}, \cite{Langlois:2002bb}.

Lately, great interests were shown in exploring some aspects of the
black strings in the context of different modified gravity models.
For example, there are several studies on the effect of the
perturbation on black strings in the de Rham-Gabadadze-Tolley (dRGT)
gravity \cite{Ponglertsakul:2018smo}, the role of the cosmological
constant in thermodynamic properties of black string
\cite{Gim:2018aix}, and the validity of entropy formula in
Einstein-Maxwell-Dilaton (EMD) theory for the black string
\cite{Setare:2018yfu}. Also, there is a tendency to extend and apply
the Anti-de Sitter/Conformal theory to different areas of physics
and since the black objects like black strings in asymptotically
Anti-de Sitter (AdS) space are noted subjects in holographic duality
viewpoint, there are several studies on this topic
\cite{Ponglertsakul:2018smo}-\cite{Nakas:2019rod}. Other current
works on black string are: The study of black string properties in
different higher dimensional theories like Lovelock theories
\cite{Giacomini:2018sho}, and the investigation of black strings in
modified gravities like Chern-Simons modified gravity (CSMG)
\cite{Cisterna:2018jsx}.

Studying how the massive and massless particles move around the
black objects is a common method to probe the gravitational fields
around them \cite{Hackmann:2009rp}. In four-dimensional space-time,
there are lots of studies that widely analyze particle motions, such
as what has investigated how the charged and neutral particles move
near the weakly magnetized Schwarzschild black hole and their
bounded trajectories \cite{Frolov:2010mi}, and the dynamics of a
charged particle in the magnetized Janis-Newman-Winicour space-time
\cite{Babar:2015kaa}. Besides, the particle's motion around a
(2+1)-dimensional BTZ (Banados, Teitelboim, Zanelli) black hole
\cite{Soroushfar:2015dfz} and geodesic equations in dilaton black
holes \cite{Soroushfar:2016yea} have studied. Other extended forms
of a Schwarzschild black hole, such as the Reissner-Nordstrom black
hole has also studied in magnetized and non-magnetized space-times
\cite{Majeed:2014kka}. It has been shown that the cosmic string
might be detected by analyzing the orbits of a test particle around
of a Kerr black hole pierced by a cosmic string
\cite{Hackmann:2010ir}, such as how the charged particles move
around the five-dimensional rotating black hole, how a magnetic
field affects this space-time explored \cite{Kaya:2007kh},
\cite{Grunau:2013oca}, how the energy of collision of two massive
particles impact in the static, rotating and magnetized black string
\cite{Tursunov:2013zha}, and the solution of geodesic equations of a
Schwarzschild black hole pierced by a cosmic string for both massive
and massless particles \cite{Hackmann:2009rp}.

Both theoretical and experimental studies show that the existence of
magnetic fields around the black holes ( due to the presence of
accretion disk's plasma \cite{Frolov:2010mi},\cite{Akiyama:2019fyp}
) would leave an impact on particle motions around the black
objects. However, in most cases while the magnetic field around the
event horizon of a black hole is not strong enough to disturb the
geometry of the black hole, it would affect the motion of charged
particles around it\footnote{These kinds of black holes are known as
"weakly magnetized" \cite{Frolov:2010mi}, \cite{Majeed:2014kka},
\cite{Znajek:1976ds}, \cite{Blandford:1977ds}, \cite{Frolov:2011ea}
${B\sim10^{4}-10^{8}\ll10^{14}}$ Gauss.}. Therefore, the magnetic
features of black holes are taken into consideration in studies
\cite{Frolov:2010mi},\cite{Majeed:2014kka},\cite{Wald:1974np}. Also,
it has been shown that the magnetic field has an important impact on
the existence of stable circular orbits in five-dimensional
space-time. Although there exist stable circular orbits around the
four-dimensional rotating black holes for massive particles, there
are no such orbits in five-dimensional one in the absence of a
magnetic field. However, in the presence of a magnetic field, there
could be such orbits for both rotating and non-rotating black holes
in five dimensions \cite{Kaya:2007kh}.

All of these points motivated us to study the massive particle
dynamics around a static black string. In this article, we first
found the effective potential of a static black string, with a
non-compact extra dimension, and the equations of motion for
time-like particles in the absence of a magnetic field. Then by
adding a magnetic field, new dynamical equations for charged
particles were found. Also, the ISCO in the dimensionless form of
dynamical equations for a massive test particle was calculated.
Then, we studied how the shape of effective potential could be
influenced by the constants of motion, especially by the new
constant of motion, we also investigated the manner in which they
influence the existence of ISCO for a massive test particle in this
metric. We calculated the escape velocity of the neutral and charged
particles in certain conditions and analyzed the effects of
different constants of motions in relation to the escape velocity
and the distance of the particle from the black string. Finally, the
analytical solutions of the equations of motion were discussed, and
some possible orbits for particles in the black string space-time
were plotted.

\section{Particles Around Black String}
In this section, we study how a massive particle moves around a
static black string space-time. This metric obtains by adding an
extra infinite spatial dimension $\omega$ to the Schwarzschild
space-time and the resulting space-time is axially symmetric.

\subsection{Metric In the Absence of Magnetic Field}
Before finding the equations of motion of a massive test particle
moving in the vicinity of a weakly magnetized black string, let us
present results for a simpler case of a magnetic field-free
condition. According to Gregory and Laflamme model, the
five-dimensional metric is \cite{Gregory:1987nb}:
\begin{equation}\label{ (1-1)}
ds^2=-F(r)dt^2+\frac{dr^2}{F(r)} +r^2(d\theta^2 +sin^2\theta
d\varphi^2)+d\omega^{2},
\end{equation}
where
\begin{equation}\label{ (1-1-1)}
F(r)=1-\dfrac{2M}{r},
\end{equation}
in which $\omega$ represents the new dimension, and $M$ is the mass
of the black string \footnote{The mass of the black string can
compute by the parameter of $\lambda$ as mass per unit length (along
the fifth axis) which compute by $ \lambda \equiv \int d^{3}x T_{00}
$
 and if the new direction is periodic
with $0 \leqslant \omega \leqslant L $, the total mass of the source
is $M=\lambda L$\cite{Lee:2006jx}.} and the gravitational constant
and the light velocity are adjusted to 1, $G=c=1$ \cite{Lee:2006jx},
\cite{Grunau:2013oca}. There is more than one independent angular
momentum in higher-dimensional axially symmetric space-times
\cite{Kunz:2014zxa}. In the case of five-dimensional space-time,
there are two independent angular momentums $L_{z}$ and $J$ related
to the test particle. These momentums are associated with the
rotation of orthogonal independent spatial planes around the two
axes. For instance, $\theta=\frac{\pi}{2}$ plane is orthogonal to
both z-axis (in Cartesian coordinate) and the fifth dimension axis.
Besides, Killing vectors are obtained as \cite{Lacquaniti:2009wc},
\cite{Frolov:2010mi}, \cite{Majeed:2014kka}, \cite{Zahrani:2013up}:
\begin{eqnarray}\label{(1-2)}
\xi _{(t)}^{\mu} \partial_{\mu}=\partial_{t},~~\xi _{(t)}^{\mu}=(1,0,0,0,0),\nonumber\\
\xi_{(\varphi)}^{\mu} \partial_{\mu}=\partial_{\varphi},~~\xi_{(\varphi)}^{\mu}=(0,0,0,1,0),\nonumber\\
\xi _{(\omega)}^{\mu}\partial_{\mu}=\partial_{\omega},~~\xi
_{(\omega)}^{\mu}=(0,0,0,0,1).
\end{eqnarray}
Therefore, there are three conserved quantities associated with the
Killing vectors, the energy( per unit mass) $\varepsilon $ of the
moving particle, the component of the angular momentum of the
particle that is aligned with z-axis $L_{z}$ and the angular
momentum $J$, which is the new constant of motion according to the
extra dimension $\omega$.
\begin{eqnarray}\label{(1-3)}
\varepsilon & \equiv &-\dfrac{P_{\mu}\xi _{(t)}^{\mu} }{m}=\dot{t}~F(r),\nonumber\\
L_{z}& \equiv &\dfrac{P_{\mu}\xi _{(\varphi)}^{\mu} }{m}=\dot{\varphi}~r^{2}~ {\sin^{2}\theta},\nonumber\\
J& \equiv &\dfrac{P_{\mu}\xi _{(\omega)}^{\mu} }{m}=\dot{\omega},
\end{eqnarray}
where $P^{\mu}=m u^{\mu}$ is the five-momentum in which $u^{\mu}$ is
the five-velocity, $m$ is the mass of the particle, and the over dot
denotes the derivative with respect to the proper time ($\tau$).

Using the invariance condition  ${g_{\mu\nu}
\dot{x}^{\mu}\dot{x}^{\nu}\equiv\epsilon}$, the constraint equation
obtains as:
\begin{equation}\label{(1-81)}
\dfrac{ds^2}{d\tau^{2}}=\epsilon\Longrightarrow \epsilon
F(r)=\dot{r}^{2}-\dot{t}^{2}~F(r)^{2}+r^{2} F(r)
\dot{\theta}^{2}+r^{2}F(r) {\sin^{2}
\theta}\dot{\varphi}^{2}+F(r)\dot{\omega}^{2}.
\end{equation}
According to Eq.~(\ref{(1-3)}) :
\begin{equation}\label{(1-4)}
\dot{r}^{2}=\varepsilon^{2}-r^{2} F(r)
\dot{\theta}^{2}-F(r)(-\epsilon+J^2+\frac{{L_{z}}^{2}}{r^{2}
{\sin^{2} \theta}}).
\end{equation}
and by defining effective potential as:
\begin{equation}\label{(1-5)}
U_{eff}=F(r)(-\epsilon+ J^2+\frac{{L_{z}}^{2}}{r^{2} {\sin^{2}
\theta}}),
\end{equation}
obtains:
\begin{equation}\label{(1-32)}
\dot{r}^{2}=\varepsilon^{2}-r^{2} F(r) \dot{\theta}^{2}- U_{eff}.
\end{equation}
where ${\epsilon=-1, 0}$ are for time-like and null geodesics
respectively. Eq.~(\ref{(1-5)})shows that the $F(r)J^2$ term is the
only difference between the effective potential in four and
five-dimensional space-times \cite{Zahrani:2013up}.

The geodesic equation (for a neutral particle) is:
\begin{equation}\label{(1-3313)}
\ddot{x}^{\mu}+\Gamma_{\nu\sigma}^{\mu}\dot{x}^{\nu}\dot{x}^{\sigma}
=0.
\end{equation}
Then, the equations of motion of time-like particles are:
\begin{equation}\label{(1-6)}
\ddot{r}=(r-3M)\dot{\theta}^{2}+(r-3M)\frac{{L_{z}}^{2}}{r^{4}{\sin^{2}
\theta}}-\frac{M}{r^{2}}(1+J^{2}),
\end{equation}
\begin{equation}\label{(1-7)}
\ddot{\theta}=-\frac{2}{r}\dot{r}\dot{\theta}+\frac{{L_{z}}^{2}\cos
{\theta}}{r^{4} {\sin^{3} \theta}}.
\end{equation}
There is no explicit relation between $J$ and $\ddot{\theta}$,
though $J$ implies its effect through $\dot{r}$ and $\dot{\theta}$.

\subsection{Presence of Magnetic Field}
Here, we study the effect of an external magnetic field on the
motion of test particles with electric charge $q$ around a black
string. Also, possible magnetic fields along each of the z-axis or
$\omega$-axis may be considered to track the test particle
trajectory. Our premise is a static, axisymmetric and homogeneous
magnetic field at the spatial infinity in the vicinity of black
string, which is directed along the z-axis. In the presence of
magnetic field the Lagrangian of the moving particle is:
\begin{equation}\label{(1-8)}
\mathfrak{L}=\dfrac{1}{2}g_{\mu\nu}\dot{x}^{\mu}\dot{x}^{\nu}+\dfrac{q}{m}
A_{\mu}\dot{x}^{\mu},
\end{equation}
where ${m}$ and ${q}$ are mass and charge of particle respectively.
The 5-potential ${A^{\mu}}$ for this metric is constructed from the
Killing vectors of the black string space-time. The Killing vectors
obey ${\square\xi^{\mu}=0}$, similar to the Maxwell equation for
4-potential in the Lorenz gauge ${A^{\mu}_{;\mu}=0}$
\cite{Frolov:2010mi}, \cite{Majeed:2014kka}, \cite{Wald:1974np},
\cite{Aliev:2002nw}:

\begin{equation}\label{(1-9)}
A^{\mu}=\dfrac{\mathfrak{B}}{2}\xi^{\mu}_{(\varphi)}-\dfrac{Q}{2m}\xi^{\mu}_{(t)}+\alpha\xi^{\mu}_{(\omega)},
\end{equation}
in which Q is the charge of black string that in our case is zero
and $\alpha$ is an arbitrary constant and since there is not any
coupling among $\omega$ and other coordinates, it could be
considered as zero too \cite{Aliev:2002nw}, and ${\mathfrak{B}}$ is
the magnetic field strength and the 2-form magnetic field, with
respect to an observer whose 5-velocity is $u_{\nu}$
\cite{McDavid:2006sq}:
\begin{equation}\label{(1-9-1)}
\mathfrak{B}^{\mu\eta}=-\dfrac{1}{2}~e^{\mu\eta\nu\lambda\sigma}~F_{\lambda\sigma}~u_{\nu},
\end{equation}
where
${e^{\mu\eta\nu\lambda\sigma}=\dfrac{\epsilon^{\mu\eta\nu\lambda\sigma}}{\sqrt{-g}},~
\epsilon_{01234}=1 ,~g=det(g_{\mu\nu})} $. The Maxwell tensor
${F_{\mu\nu}}$ is defined as \cite{McDavid:2006sq}:
\begin{equation}\label{(1-10)}
F_{\mu\nu}= A_{\nu,\mu}- A_{\mu,\nu}= A_{\nu;\mu}- A_{\mu;\nu}.
\end{equation}
By considering Eq.~(\ref{(1-2)}) and Eq.~(\ref{(1-9)}) one can see
that the 5-potential ${A^{\mu}}$ is invariant with respect to the
Killing vector isometries \cite{McDavid:2006sq}:
\begin{equation}\label{(1-999)}
(L_{\xi}A)_{\mu}=A_{\mu,\nu} \xi^{\nu}+A_{\nu} \xi^{\nu}_{,\mu}=0.
\end{equation}
The conserved quantities associated with these symmetries are
obtained as \cite{Frolov:2010mi}, \cite{Majeed:2014kka}:
\begin{eqnarray}
\label{(1-11)}
\varepsilon & \equiv &-\dfrac{P_{\mu}\xi _{(t)}^{\mu} }{m}=\dot{t}~F(r),\nonumber\\
L_{z} &\equiv & \dfrac{P_{\mu}\xi _{(\varphi)}^{\mu} }{m}=(\dot{\varphi}+B)~r^{2}~ {\sin^{2}\theta},\nonumber\\
J &\equiv & \dfrac{P_{\mu}\xi _{(\omega)}^{\mu} }{m}=\dot{\omega},
\end{eqnarray}
where ${P^{\mu}=m u^{\mu}+qA^{\mu}}$ ,and $q$ is the charge of the
particle and:
\begin{equation}\label{(1-16)}
B=\dfrac{q\mathfrak{B}}{2m}.
\end{equation}

By using  the normalization condition, and similar to
Eq.~(\ref{(1-81)}), we obtain the effective potential for the
charged time-like particle in the presence of a magnetic field as:
\begin{equation}\label{(1-99)}
\dot{r}^{2}=\varepsilon^{2}-r^{2} F
\dot{\theta}^{2}-F(r)[-\epsilon+J^{2}+r^{2}\sin
^{2}\theta(\dfrac{L_{z}}{r^{2} \sin^{2} \theta}-B)^{2}],
\end{equation}
\begin{equation}\label{(1-12)}
U_{eff}=F(r)[-\epsilon+J^{2}+r^{2}\sin
^{2}\theta(\dfrac{L_{z}}{r^{2} \sin^{2} \theta}-B)^{2}].
\end{equation}
As we know, the equation of motion of the charged particle in an
electromagnetic field  ${F_{\mu\nu}}$ is \cite{Frolov:2010mi},
\cite{Majeed:2014kka}:
\begin{equation}\label{(1-13)}
\ddot{x}^{\mu}+\Gamma_{\nu\sigma}^{\mu}\dot{x}^{\nu}\dot{x}^{\sigma}
=\dfrac{q}{m}F_{\alpha}^{\mu}\dot{x}^{\alpha}.
\end{equation}
Therefore, the dynamical equations in black string metric are:
\begin{equation}\label{(1-14)}
\ddot{r}=(r-3M)\dot{\theta}^{2}+(r-3M)\dfrac{L_{z}^{2}}{r^{4}
\sin^{2} \theta}+ \dfrac{2M}{r^{2}}B L_{z}-(r-M)\sin^{2} \theta
B^{2}-\dfrac{M}{r^{2}}(1+J^{2}),
\end{equation}
\begin{equation}\label{(1-15)}
\ddot{\theta}=-\dfrac{2}{r}\dot{r}\dot{\theta}+\dfrac{L_{z}^{2}\cos
\theta}{r^{4} \sin^{3} \theta}-B^{2}\sin{\theta} \cos{\theta} .
\end{equation}
Similar to the magnetic field free case, the $\ddot{\theta}$ is
independent of the new angular momentum $J$.

\section{Dimensionless Dynamical Equations}
For more simplification, one can rewrite the dynamical equations in
dimensionless form, by using these definitions \cite{Frolov:2010mi},
\cite{Majeed:2014kka}, \cite{Zahrani:2013up}:
\begin{equation}\label{(2-1)}
\sigma=\dfrac{\tau}{2M} , ~~\rho=\dfrac{r}{2M} ,
~~l_{z}=\dfrac{L_{z}}{2M} , ~~\Omega=\dfrac{\omega}{2M} , ~~b=2 B M
, ~~ \acute{m}=\dfrac{m}{2M} .
\end{equation}
So the Eq.~(\ref{(1-14)}) and Eq.~(\ref{(1-15)}) are expressed as
follows:
\begin{equation}\label{(2-2)}
\dfrac{d^{2}\theta}{d\sigma^{2}}=-\dfrac{2}{\rho}\dfrac{d\rho}
{d\sigma}\dfrac{d\theta}{d\sigma}+\dfrac{l_{z}^{2}\cos\theta}
{\rho^{4}\sin ^{3}\theta}-b^{2}\sin \theta \cos \theta,
\end{equation}
\begin{equation}\label{(2-3)}
\dfrac{d^{2}\rho}{d\sigma^{2}}=\dfrac{1}{2}(2\rho
-3)({\dfrac{d\theta}{d\sigma}})^{2}+\dfrac{2bl_{z}-1}{2\rho^{2}}+\dfrac{l_{z}^{2}(2\rho
-3)}{2\rho^{4}\sin ^{2}\theta}-\dfrac{b^{2}}{2}(2\rho -1)\sin
^{2}\theta-\dfrac{J^{2}}{2\rho^{2}},
\end{equation}
and the Eq.~(\ref{(1-99)}) for time-like geodesics (${\epsilon=-1}$)
changes to:
\begin{equation}\label{(2-4)}
{(\dfrac{d\rho}{d\sigma})}^{2}=\varepsilon^{2}-\rho^{2}(1-\dfrac{1}{\rho})
{(\dfrac{d\theta}{d\sigma})}^{2}-U_{eff},
\end{equation}
in which:
\begin{equation}\label{(2-5)}
U_{eff}=(1-\dfrac{1}{\rho})[1+J^{2}+\dfrac{({l_{z}-b\rho^{2}\sin
^{2}\theta})^{2}} {\rho^{2}\sin ^{2}\theta}].
\end{equation}
Extending some four-dimensional cases in the planetary motion of
test particle studies \cite{Frolov:2010mi}, \cite{Majeed:2014kka},
\cite{Zahrani:2013up} to five-dimensional analysis, we choose
$\theta=\dfrac{\pi}{2}$ plane as our test particle's trajectory
plane.

\subsection{The Innermost Stable Circular Orbit}
According to Eq.~(\ref{(2-5)}) and as it is shown in the next
section the angular momentum has the role of a defining parameter of
the shape of potential \cite{Jefremov:2015gza}. Besides, like its
important role in the shape of potential in the Schwarzschild black
hole \cite{Jefremov:2015gza}, \cite{Frolov:2010mi}, the amount of
${l_{z}}$ determines if the 5-dimensional effective potential has
both a minimum and a maximum, only one extremum, or no extremums at
all. So, to find the innermost stable circular orbits (ISCO), we not
only need to find ${\rho_{ISCO}}$, but also ${l_{z ISCO}}$. In fact,
we consider the series of mentioned conditions of circular motion in
ref. \cite{Jefremov:2015gza}, to find ISCO parameters:

(i) Firstly, in circular motion the radial velocity must be zero so:
$ {\dfrac{d\rho}{d\tau}=0} $.

(ii) Secondly, the acceleration along the radial coordinate should
be zero:  $ {\dfrac{d^{2}\rho}{d\tau^{2}}=0} $.

(iii) Thirdly, the second derivative of effective potential must be
zero in an inflection point.

All of these conditions must be satisfied, simultaneously. Our
equations of energy and effective potential in the presence of
magnetic field are Eq.~(\ref{(2-4)}) and Eq.~(\ref{(2-5)}). The
first condition (in the $ {\theta=\dfrac{\pi}{2}} $ plane) leads to
$ {\varepsilon^{2}=U_{eff}} $ and the second one leads to:
\begin{equation}\label{(2-6)}
U_{,\rho}=l_{z}^{2}(3-2\rho)- 2 b
l_{z}\rho^{2}+\rho^{2}[1+J^{2}+b^{2} \rho^{2}(2\rho-1)]=0,
\end{equation}
where ${"_{,\rho}"=\dfrac{\partial}{\partial\rho}}$ , and the third
one is:
\begin{equation}\label{(2-7)}
U_{,\rho\rho}=2\left[\rho^{2}(b^{2}\rho^{3}-J^{2}-1)+2bl_{z}\rho^{2}+3l_{z}^{2}(\rho-2)\right]=0.
\end{equation}
where ${"_{,\rho\rho}"=\dfrac{\partial^{2}}{\partial\rho^{2}}}$. In
general, finding the exact inflection point of Eq.~(\ref{(2-6)}) is
hard and it should be solved numerically \cite{Jefremov:2015gza}.
Therefore, we choose the perturbation method that has been suggested
in \cite{Jefremov:2015gza}. We presume that although the magnetic
field near the black string can disturb it, yet its magnitude is
small enough ${(b \ll 1)}$ to use the perturbation method. ISCO
parameters in the absence of magnetic field are ${\rho_{0}}$,
${l_{z0}}$ and ${\varepsilon_{0}}$:
\begin{equation}\label{(2-9)}
\rho_{0}=3,
\end{equation}
\begin{equation}\label{(2-10)}
l_{z0}=\pm\sqrt{3(1+J^{2})},
\end{equation}
\begin{equation}\label{(2-11)}
\varepsilon_{0}=\dfrac{2\sqrt{2}}{3}\sqrt{1+J^{2}}.
\end{equation}
This shows that, among these three parameters, only ${l_{z0}}$ and
${\varepsilon_{0}}$ are related to the $J$. By choosing:
\begin{equation}\label{(2-12)}
\rho=\rho_{0}+b \rho_{1}+b^{2}\rho_{2},
\end{equation}
\begin{equation}\label{(2-13)}
l_{z}=l_{z0}+b l_{z1}+b^{2}l_{z2},
\end{equation}
\begin{equation}\label{(2-14)}
\varepsilon = \varepsilon_{0}+b
\varepsilon_{1}+b^{2}\varepsilon_{2},
\end{equation}
and substituting them in $ {U_{,\rho}=0}$, $ {U_{,\rho\rho}=0}$ and
neglecting the third order of the perturbation, the results are:
\begin{equation}\label{(2-15)}
\rho_{1}=0,~~l_{z1}=-3,~~\varepsilon_{1}=\mp\dfrac{\sqrt{3}}{3} ,
\end{equation}

\begin{equation}\label{(2-1515)}
\rho_{2}=\dfrac{-216}{1+J^{2}},
~~l_{z2}=\dfrac{\mp72}{\sqrt{3(1+J^{2})}},
~~\varepsilon_{2}=\dfrac{8}{\sqrt{2(1+J^{2})}},
\end{equation}

so:
\begin{equation}\label{(2-16)}
\rho = 3-\dfrac{216}{1+J^{2}}b^{2},
\end{equation}
\begin{equation}\label{(2-17)}
l_{z}=\pm\sqrt{3(1+J^{2})}-3b\mp\dfrac{72}{\sqrt{3(1+J^{2})}}b^{2},
\end{equation}
\begin{equation}\label{(2-18)}
\varepsilon =
\dfrac{2\sqrt{2}}{3}\sqrt{1+J^{2}}\mp\dfrac{\sqrt{3}}{3}
b+\dfrac{8}{\sqrt{2(1+J^{2})}} b^{2}.
\end{equation}

Up to the first order of perturbation, these equations show that the
radius of ISCO in Schwarzschild black hole is similar to that of the
black string, and the fifth dimension has its effect on ${l_{z}}$
and ${\varepsilon}$ via new angular momentum $J$. Here we have three
parameters $l_{z}$, $J$, $\varepsilon$ which are constants of
motion. We assume ${J}$ as our degree of freedom parameter and tuned
it in a way that helps us find the other two parameters. Now, in
general condition and similar to \cite{Frolov:2010mi}, by adding
Eq.~(\ref{(2-6)}) and Eq.~(\ref{(2-7)}) we can eliminate ${J }$ and
find ${l_{z} }$ in terms of ${b }$:
\begin{equation}\label{(2-19)}
l_{z\pm}=\pm b
\rho^{2}\left[\dfrac{1-3\rho}{\rho-3}\right]^{\dfrac{1}{2}},
\end{equation}
in which, ${l_{z-}}$ shows that the Lorentz force is attractive and
${l_{z+}}$ how the Lorentz force is repulsive \cite{Frolov:2010mi},
\cite{Babar:2015kaa}, \cite{Zahrani:2013up}. Now by substituting
Eq.~(\ref{(2-19)}) into Eq.~(\ref{(2-7)}) we can find parameter ${b
}$ in terms of ${\rho }$ and  ${J }$:
\begin{equation}\label{(2-20)}
b=\dfrac{\sqrt{2
(3-\rho)(1+J^{2})}}{2\rho\left[4\rho^{2}-9\rho+3\pm\sqrt{(3-\rho)(3
\rho -1)}\right]^{\dfrac{1}{2}}}.
\end{equation}
It shows that the new constant of motion, related to the fifth
dimension, is not separable from both ${b }$ and ${l_{z} }$.
Comparing Eq.(\ref{(2-19)}) and Eq.(\ref{(2-17)}) one can see that
adding a fifth dimension shows its effects on ${l_{z}}$ directly
Eq.(\ref{(2-17)}) or via  ${b }$ in Eq.(\ref{(2-19)}). Now by
putting Eq.(\ref{(2-20)}) in Eq.(\ref{(2-19)}), ${l_{z}}$ in terms
of ${\rho }$ will be obtained:
\begin{equation}\label{(2-21)}
l_{z}=\pm\dfrac{\rho\sqrt{(3 \rho
-1)(1+J^{2})}}{\sqrt{2}\left[4\rho^{2}-9\rho +3\pm\sqrt{(3-\rho)(3
\rho -1)}\right]^{\dfrac{1}{2}}}.
\end{equation}
Eq.(\ref{(2-20)}) and Eq.(\ref{(2-21)}) show the combinations among
parameters $b$, $l_{z}$ and $J$. Also, in the case of $J=0$, these
results are in accordance with the four-dimensional black hole
results which were already obtained in \cite{Frolov:2010mi},
\cite{Zahrani:2013up}.

Since we have only two constraint equations ($U_{,\rho\rho}=0$,
$U_{,\rho}=0$), we cannot express their relations to $\rho $
separately. So, we have to fix one of them ($b$, $l_{z}$, $J$).
According to our decision for choosing a small magnetic field, we
chose ${b=0.5 }$ (fixed).

\section{Effective Potential}
\begin{figure}[h]
\centering \subfigure[]{
\includegraphics[width=0.5\textwidth]{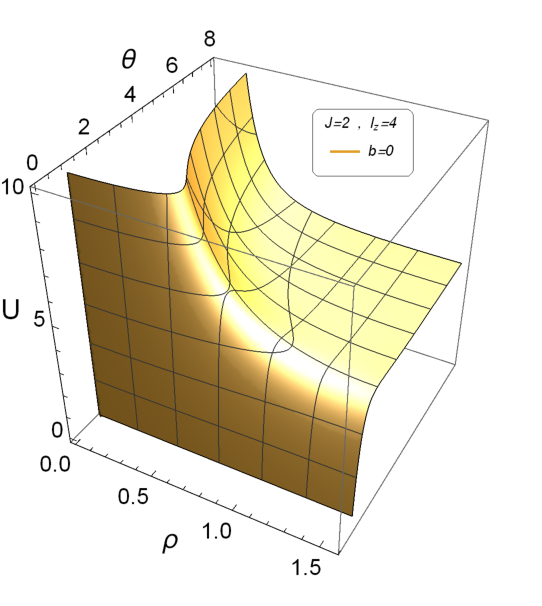}\label{ae1}
        }
\subfigure[]{
\includegraphics[width=0.5\textwidth]{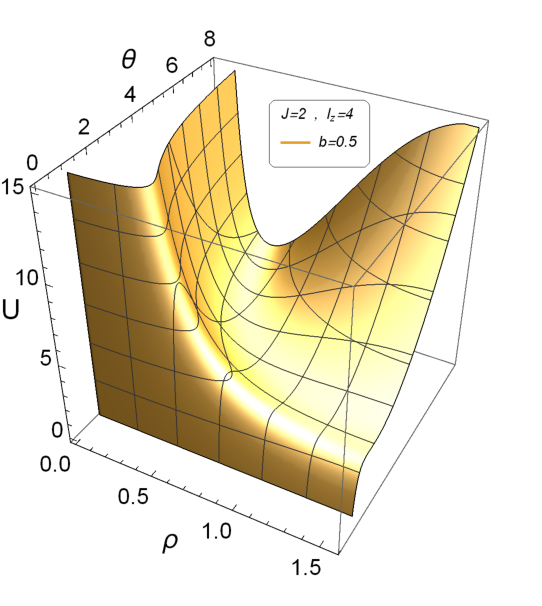}\label{ae2}
        }
\caption{Effective Potential versus $\rho$ and ${\theta}$ in $J=2$
and $l_{z}=4$ for (a) $b=0$, (b) $b=0.5$.} \label{fe11}
\end{figure}

\begin{figure}[h]
\centering \subfigure[]{
\includegraphics[width=0.9\textwidth]{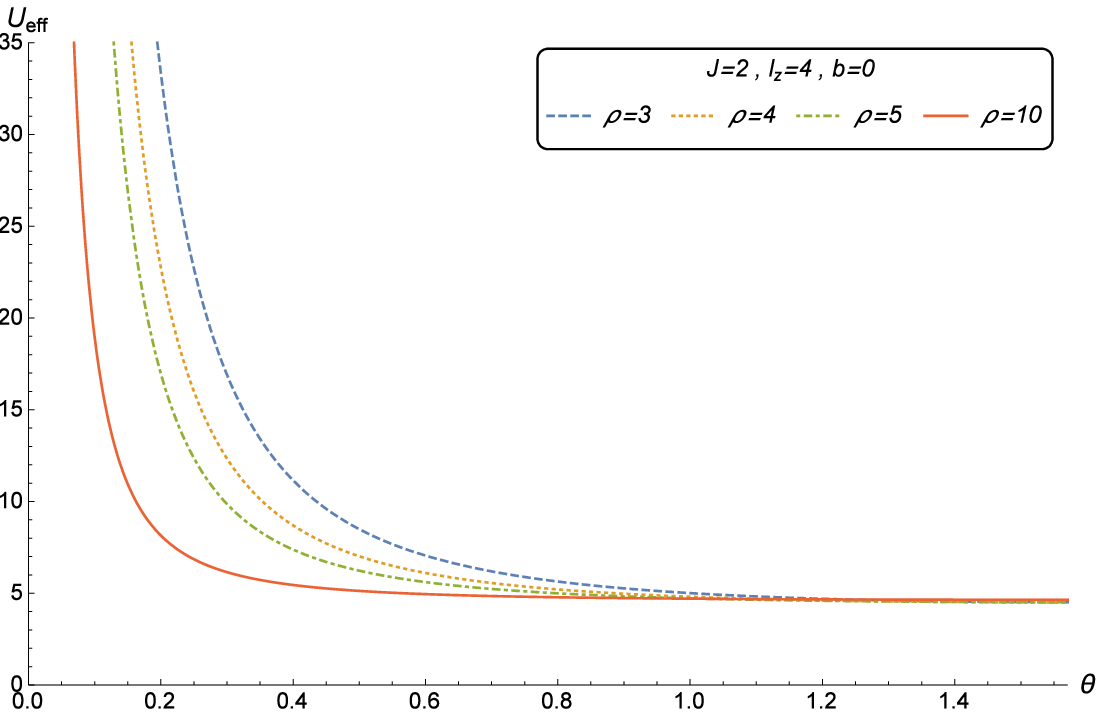}\label{ae3}
        }
\subfigure[]{
\includegraphics[width=0.9\textwidth]{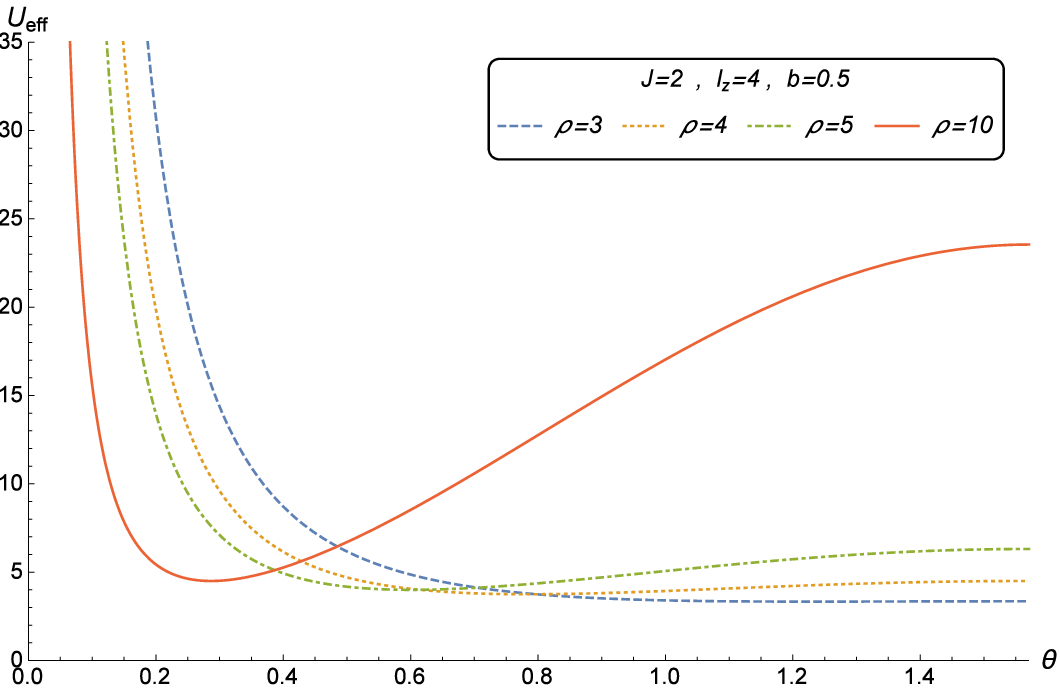}\label{ae4}
        }
\caption{Effective Potential versus ${\theta}$ }
% in $J=2$ and $l_{z}=4$
%for different $\rho$, (a) $b=0$, $\rho=3$ (blue dashed), $\rho=4$ (orange dotted),
%$\rho=5$ (green dotdashed), $\rho=10$ (red line), (b) $b=0.5$, $\rho=3$ (blue dashed),
%$\rho=4$ (orange dotted), $\rho=5$ (green dotdashed), $\rho=10$ (red line).}
 \label{fe2}
\end{figure}

\begin{figure}[h]
\centering \subfigure[]{
\includegraphics[width=0.9\textwidth]{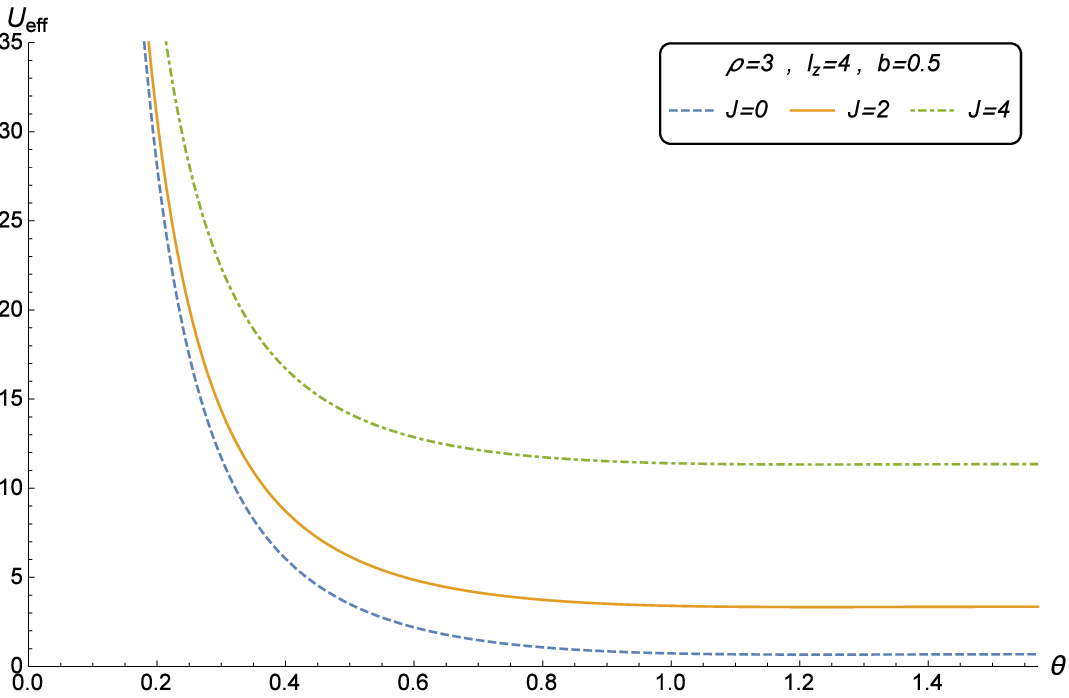}\label{ae5}
        }
\caption{Effective Potential versus ${\theta}$ in $\rho=3$, $b=0.5$
and $l_{z}=4$ for $J=0$ (blue dashed), $J=2$ (orange line), $J=4$
(green dot dashed).} \label{fe3}
\end{figure}

To study the effective potential of the black string and its
properties, we consider that the strength of the magnetic field and
the mass of the black string are fixed. Thus, based on
Eq.(\ref{(2-5)}), the effective potential depends on ${l_{z}}$,
${\rho}$, ${\theta}$ and ${J}$. The effective potential versus
${\rho}$ and ${\theta}$ is shown in Fig.\ref{fe11}. Fig.\ref{fe2}
displays the relation between effective potential versus ${\theta}$
for different ${\rho}$, in the presence and absence of a magnetic
field. It depicts that even the presence of a magnetic field with
small magnitude, makes a barrier in ${\theta=\dfrac{\pi}{2}}$,
especially for larger ${\rho}$. As it shows, in the absence of a
magnetic field and in ${\theta}=0$, there is a potential wall;
while, $U_{eff}$ is constant around the ${\theta}=\dfrac{\pi}{2}$
for a specified distance. However, in the presence of magnetic field
and especially for larger ${\rho}$, there is a potential barrier
around ${\theta}=\dfrac{\pi}{2}$. The impact of ${J}$ on the
effective potential in different ${\theta}$ is shown in
Fig.\ref{fe3}.

\begin{figure}[h]
\centering \subfigure[]{
\includegraphics[width=0.9\textwidth]{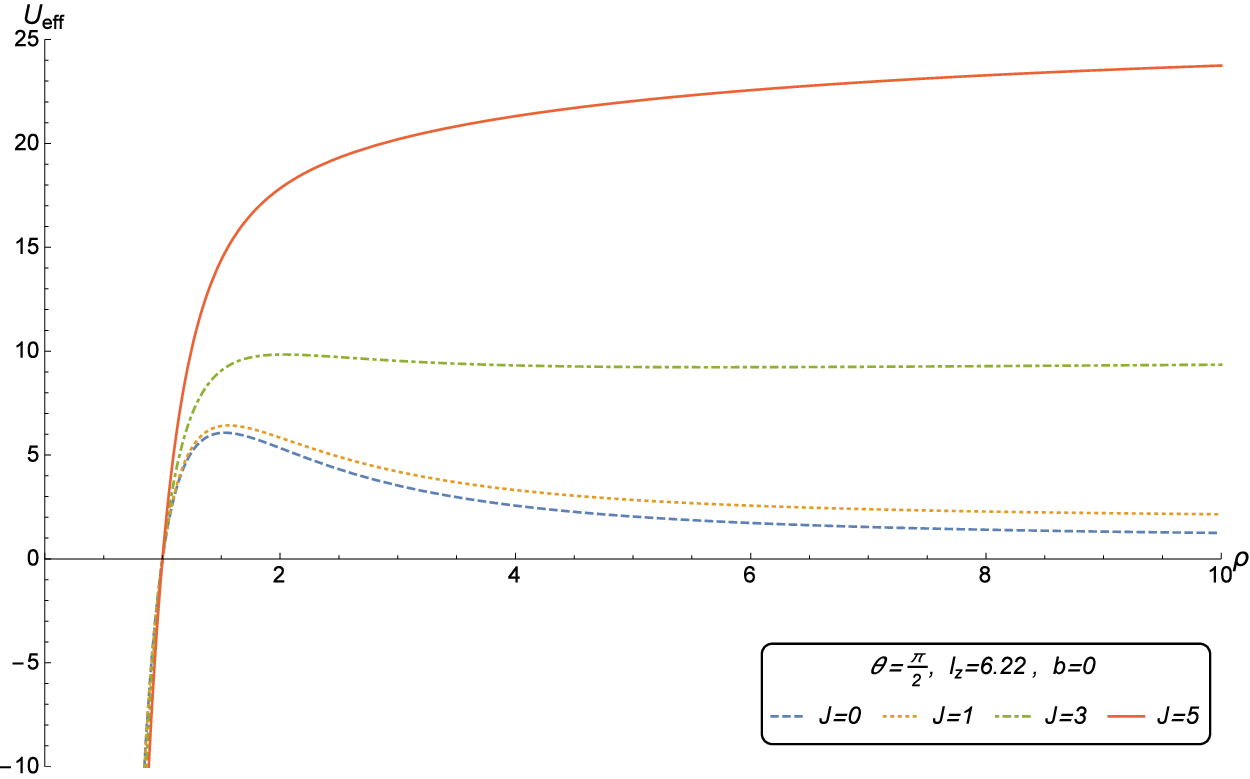}\label{a1}
        }
\subfigure[]{
\includegraphics[width=0.9\textwidth]{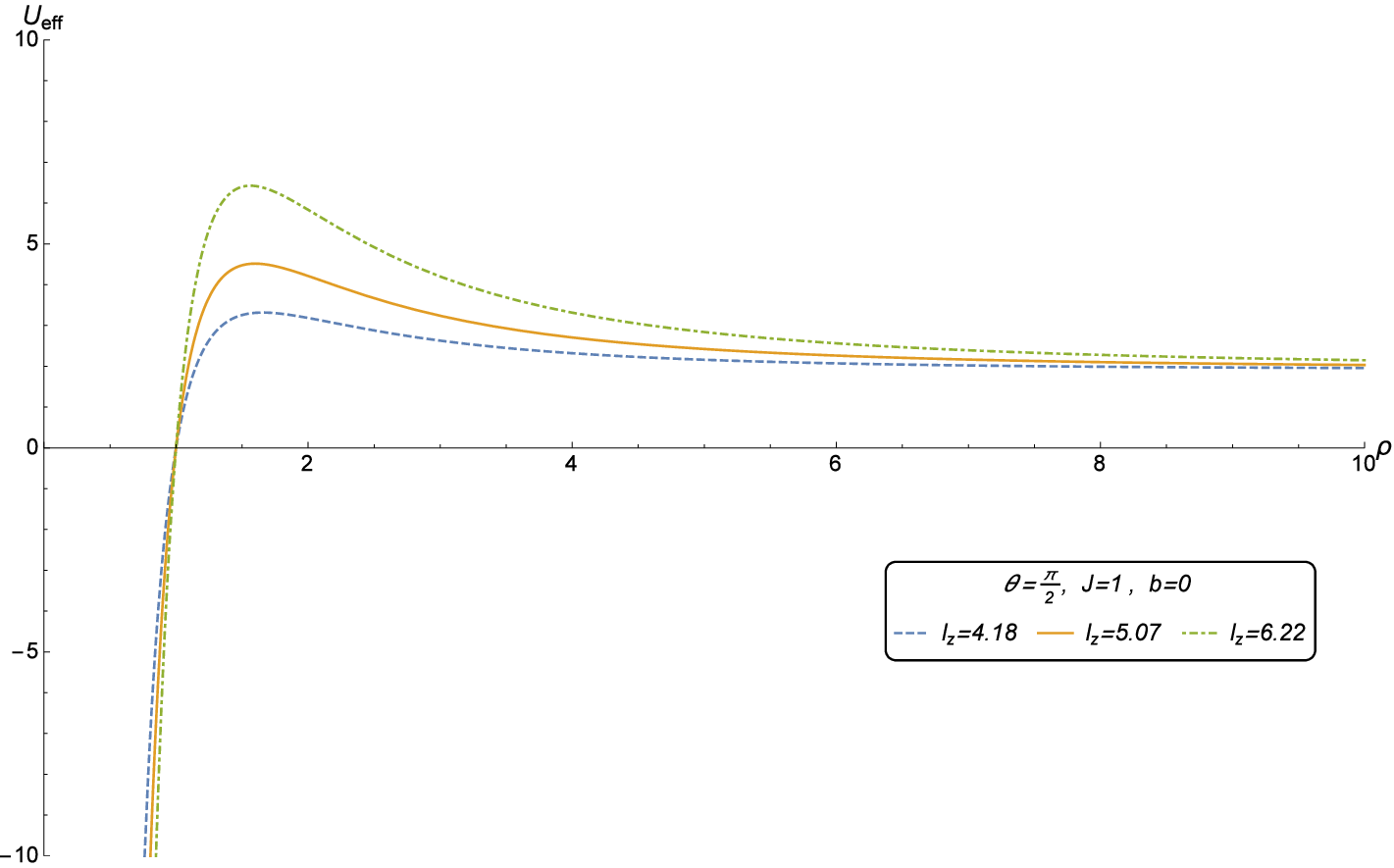}\label{a2}

        }
\caption{Effective Potential versus $\rho$ in
${\theta=\dfrac{\pi}{2}}$: (a) $b=0$, $l_{z}=6.22$ for $J=0$ (blue
dashed), $J=1$ (orange dotted), $J=3$ (green dotdashed), $J=5$ (red
line), (b) $b=0$, $J=1$ for $l_{z}=4.18$ (blue dashed), $l_{z}=5.07$
(orange line), $l_{z}=6.22$ (green dotdashed)}
 \label{f1}
\end{figure}

\begin{figure}[h]
\centering \subfigure[]{
\includegraphics[width=0.9\textwidth]{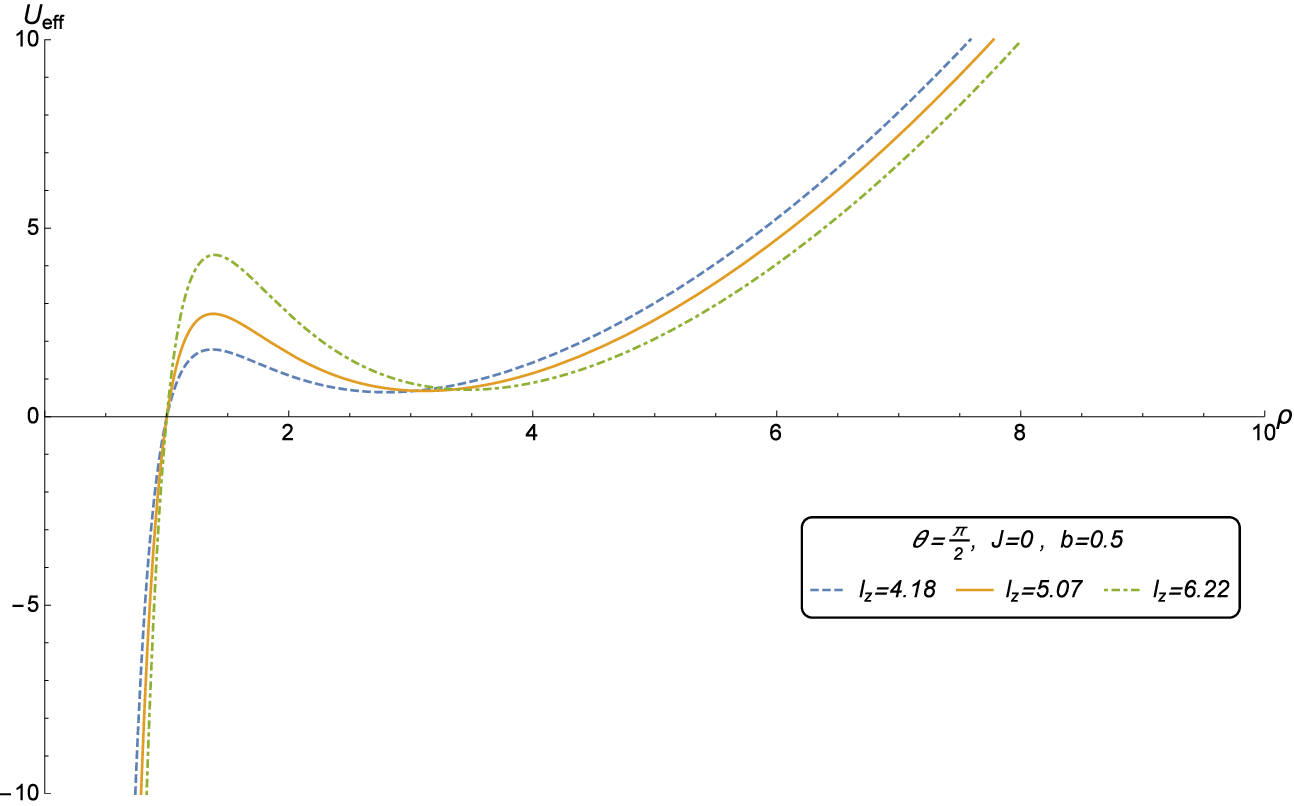}\label{a3}
        }
\subfigure[]{
\includegraphics[width=0.9\textwidth]{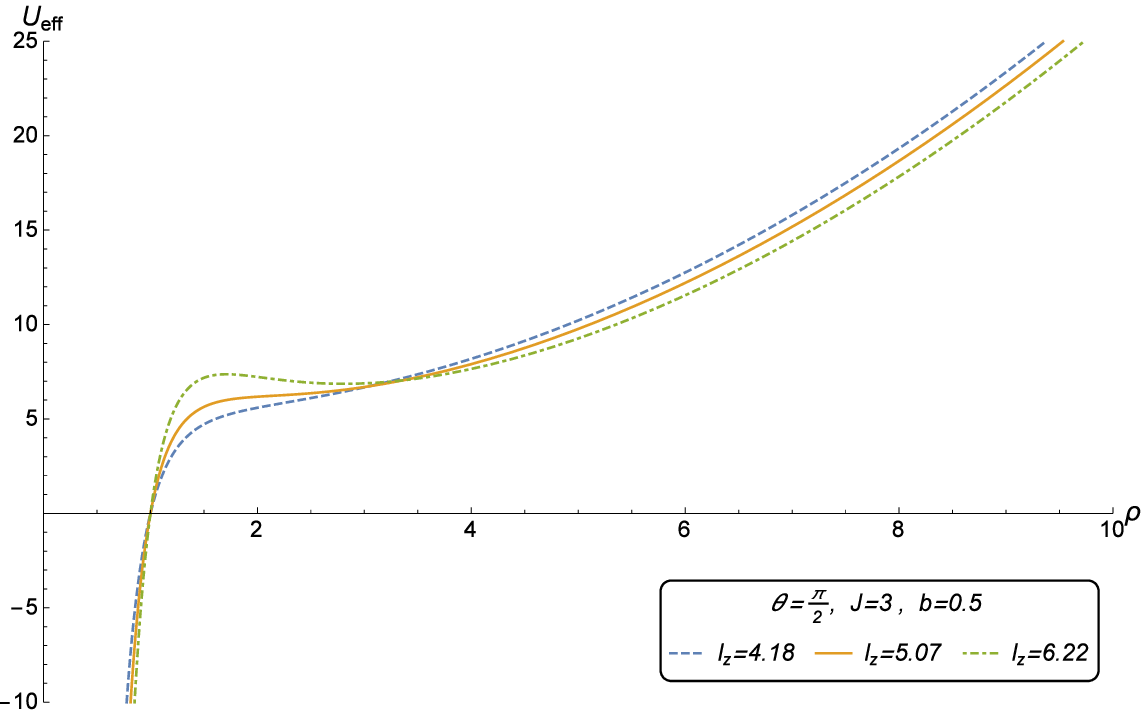}\label{a4}
      }
\caption{Effective Potential versus $\rho$ in
${\theta=\dfrac{\pi}{2}}$:(a) $J=0$, $b=0.5$ for $l_{z}=4.18$ (blue
dashed), $l_{z}=5.07$ (orange line), $l_{z}=6.22$ (green
dotdashed),(b) $J=3$, $b=0.5$ for $l_{z}=4.18$ (blue dashed),
$l_{z}=5.07$ (orange line), $l_{z}=6.22$ (green dotdashed). }
 \label{f12020}
\end{figure}

\begin{figure}[h]
\centering \subfigure[]{
\includegraphics[width=0.9\textwidth]{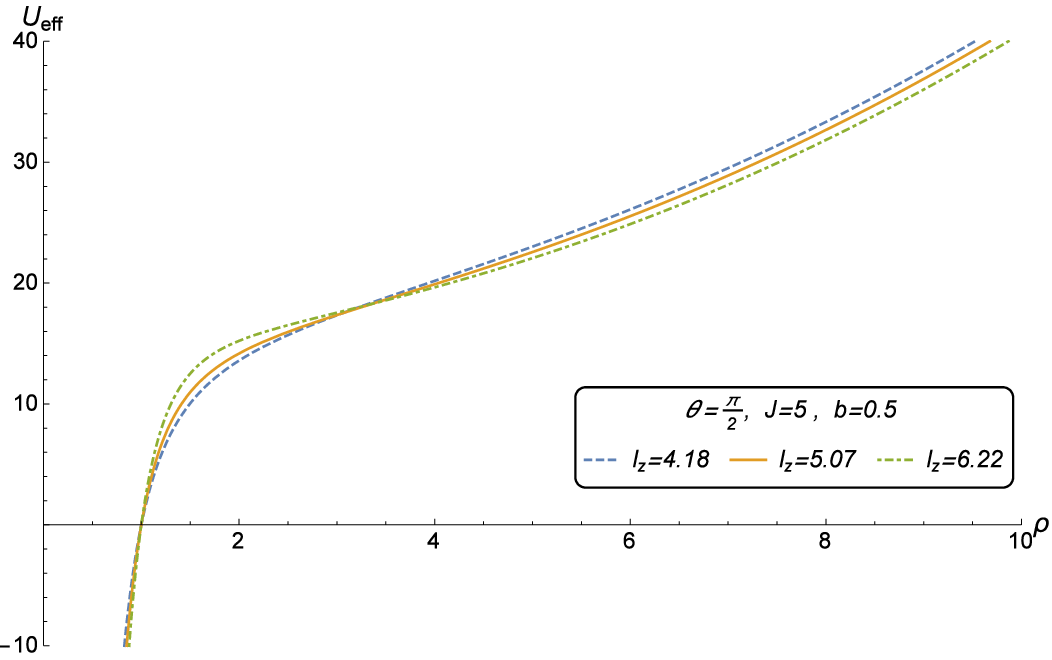}\label{a5}
        }

\subfigure[]{
\includegraphics[width=0.9\textwidth]{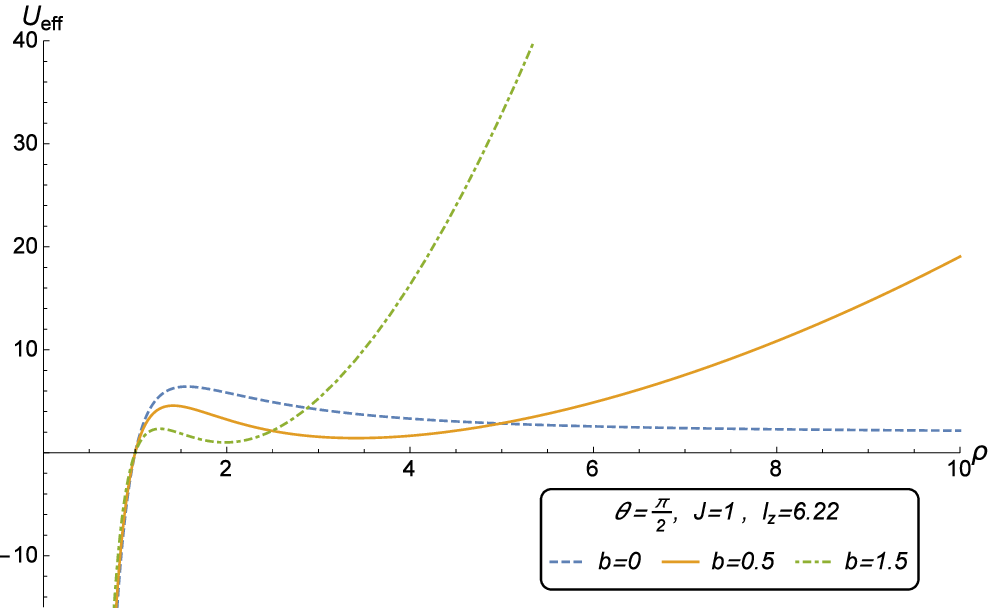}\label{a6}
}

\caption{Effective Potential versus $\rho$ in
${\theta=\dfrac{\pi}{2}}$:(a) $J=5$, $b=0.5$ for $l_{z}=4.18$ (blue
dashed), $l_{z}=5.07$ (orange line), $l_{z}=6.22$ (green dotdashed),
(b) $J=1$, $l_{z}=6.22$ for $b=0$ (blue dashed), $b=0.5$ (orange
line), $b=1.5$ (green dotdashed).}

\label{f120202}
\end{figure}

Fig.\ref{f1}, Fig.\ref{f12020}, and Fig.\ref{f120202} demonstrate
the profile of effective potential versus ${\rho}$ in
${\theta=\dfrac{\pi}{2}}$ plane for different ${J}$, ${b }$, and
${l_{z}}$. As it shows here and from Fig.\ref{ae4}, one can see that
the presence of a magnetic field has a noticeable effect on the
shape of potential especially in ${\theta}=\dfrac{\pi}{2}$. The
magnitude of ${J}$ has an important effect on it, as well.
Fig.\ref{a1} shows that, for a fixed ${l_{z}=6.22}$, in the absence
of a magnetic field, there is no stable circular orbit, which is
similar to a five-dimensional black hole in the absence of a
magnetic field as mentioned before \cite{Kaya:2007kh}, and by
increasing ${J }$ one can see how the shape of potential changes to
a slight line out of the event horizon. According to
Eq.(\ref{(2-5)}), the magnitude of potential in larger distances is
constant and equal to $(1+J^{2})$ that shows the effects of the
extra dimension on the potential in the absence of a magnetic field.
Fig.\ref{a2} illustrates the effects of the changes of ${l_{z}}$ on
the shape of the potential, in the absence of a magnetic field and a
fixed ${J=1}$.

In Fig.\ref{a3}, Fig.\ref{a4} and Fig.\ref{a5}, comparing the green
lines (with similar ${l_{z}}$) one can see that by increasing the
$J$ for fixed $b=0.5$, the shape of potential changes, the distance
between maximum and minimum of potential decreases, and the saddle
point disappears. This shows the effect of extra dimension (via $J$)
on the chance of existence of ISCO. The more the magnitude of $J$,
the less the chance of existence of ISCO. In addition, Fig.\ref{a5}
displays that for the large values of $J$, $J=5$ as a sample case,
even with a different $l_{z}$, there is no stable circular orbits.
But, as it is shown in Fig.\ref{a6}, for the proper amount of $J$
and $l_{z}$ in $b\neq0$, the potential of black string has ISCO
(inflection point in potential). As it is obvious from Fig.\ref{a1},
in the absence of a magnetic field and in large distances from black
string ($\rho\rightarrow\infty$) the effective potential is almost
constant and its magnitude depends on $J$. According to
Eq.(\ref{(2-5)}), this magnitude is ($1+J^{2}$). However, in the
presence of the magnetic field, Fig.\ref{f12020} and
Fig.\ref{f120202} show a divergent behaviour of the effective
potential in large distances ($\rho\rightarrow\infty \Longrightarrow
U_{eff}\rightarrow\infty$). In both cases (in the absence and
presence of the magnetic field), the effective potential is zero at
the event horizon ($\rho=1$), which is obvious from
Eq.(\ref{(2-5)}). Besides, the calculated radius of ISCO
\footnote{In the presence of the magnetic field by perturbation
method up to first order}($\rho_{0}=3 $), shows itself in
Fig.\ref{f120202} as an inflection point in the effective potential.

\section{effective force}
One could compute the effective force related to the effective
potential by:
\begin{equation}\label{(f-1)}
\vec{F}_{eff}=-\vec{\nabla} U_{eff},
\end{equation}
So:
\begin{equation}\label{(f-2)}
F_{\rho}=\dfrac{l_{z}^{2}(2\rho -3)}{\rho^{4}\sin^{2}
\theta}+\dfrac{2bl_{z}}{\rho^{2}}-\dfrac{1+J^{2}}{\rho^{2}}
-b^{2}(2\rho-1)\sin^{2} \theta ,
\end{equation}
and
\begin{equation}\label{(f-3)}
F_{\theta}=-(\rho-1) b^{2} \sin (2\theta)+\dfrac{2 l_{z}^{2}\cot
\theta}{\rho^{4}\sin^{2} \theta}(\rho-1).
\end{equation}

\begin{figure}[h]
\centering \subfigure[ ]{
\includegraphics[width=0.9\textwidth]{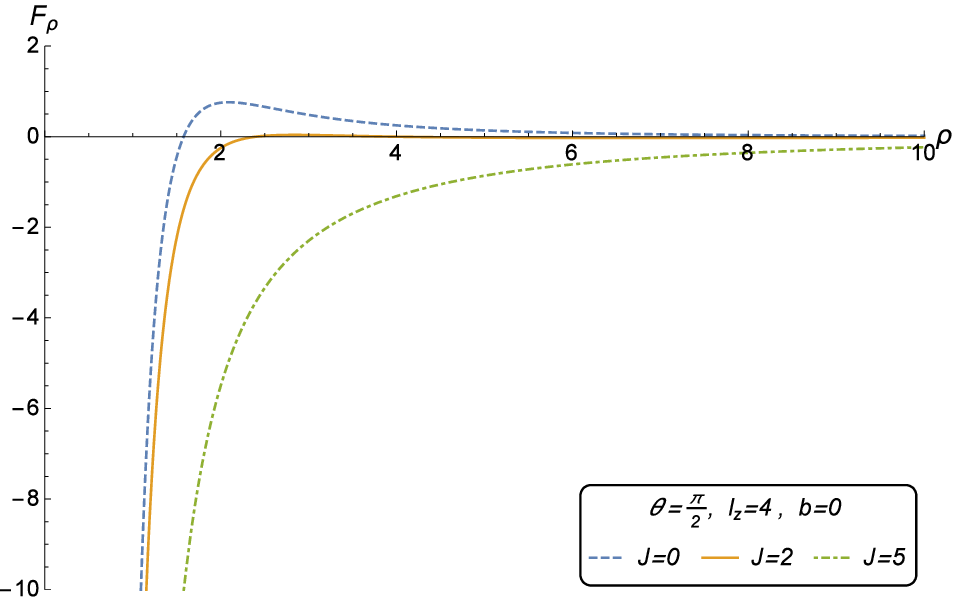}\label{rho12}
   }
\subfigure[ ]{
\includegraphics[width=0.9\textwidth]{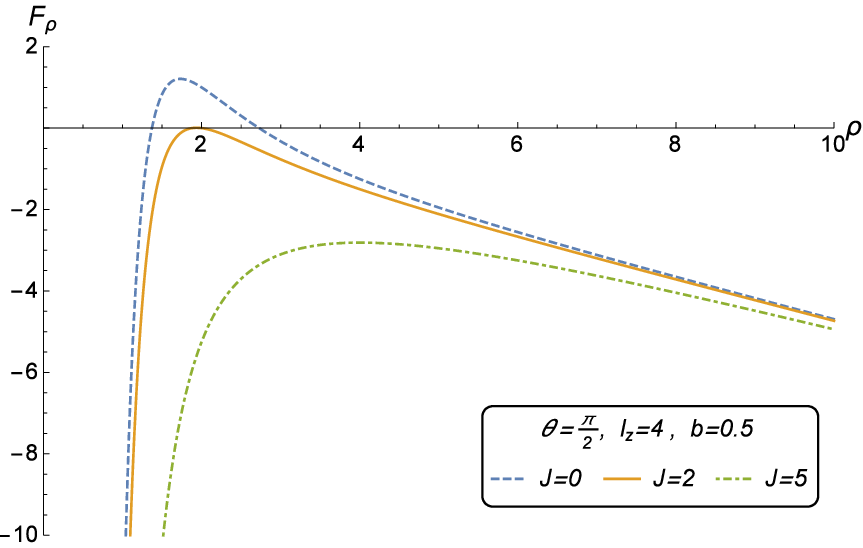}\label{rho1}
   }
\caption{Effective Force $F_{\rho}$ versus $\rho$ in $l_{z}=4$: (a)
for $b=0$, $\theta=\dfrac{\pi}{2}$ and for different $J$, $J=0$
(blue dashed), $J=2$ (orange line), $J=5$ (green dotdashed), (b) for
$b=0.5$, $\theta=\dfrac{\pi}{2}$ and for different $J$, $J=0$ (blue
dashed), $J=2$ (orange line), $J=5$ (green dotdashed).}
\label{frho1}
\end{figure}

\begin{figure}[h]
\centering \subfigure[ ]{
\includegraphics[width=0.9\textwidth]{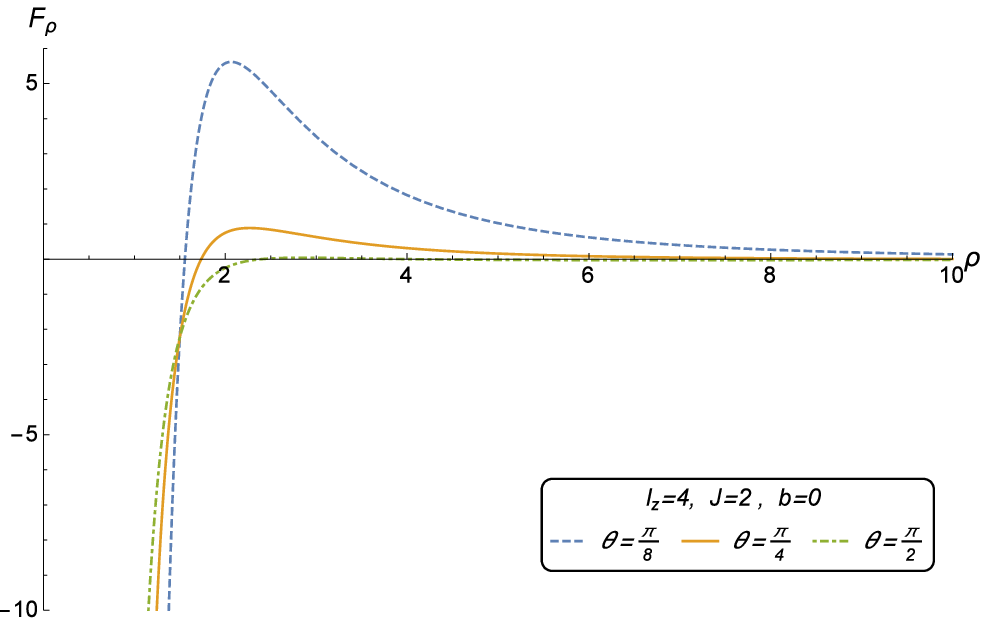}\label{rho22}
   }
\subfigure[ ]{
\includegraphics[width=0.9\textwidth]{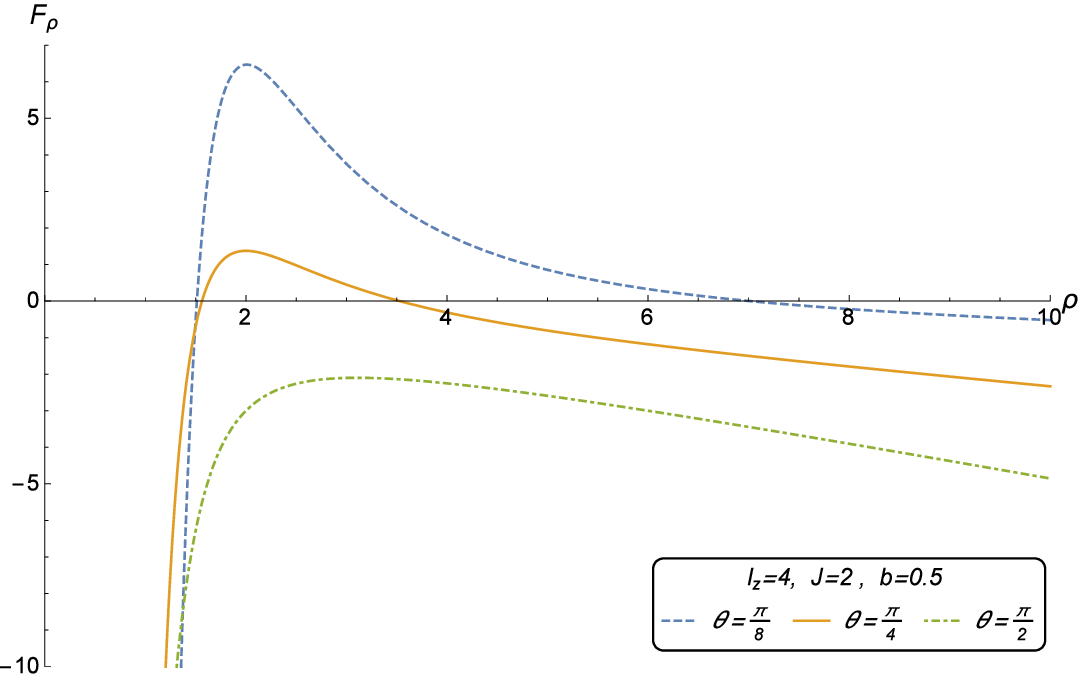}\label{rho2}
   }
\caption{Effective Force $F_{\rho}$ versus $\rho$in $l_{z}=4$: (a)
for $b=0$ in different $\theta$, $\theta=\dfrac{\pi}{8}$ (blue
dashed), $\theta=\dfrac{\pi}{4}$ (orange line),
$\theta=\dfrac{\pi}{2}$ (green dotdashed), (b) for $b=0.5$ in
different $\theta$, $\theta=\dfrac{\pi}{8}$ (blue dashed),
$\theta=\dfrac{\pi}{4}$ (orange line), $\theta=\dfrac{\pi}{2}$
(green dotdashed).}
 \label{frho}
\end{figure}

In these results, the new constant of motion only appears in
$F_{\rho}$ and has no effects on $F_{\theta}$. Also, the new
constant of motion $J$ does not couple with the magnetic field.
Eq.(\ref{(f-2)}) shows that for the fixed positive $b$ and $l_{z}$,
and for distances $\rho\geqslant 1.5$, the first and the second
terms in $F_{\rho}$ are positive. It indicates that $l_{z}$ plays a
repulsive role in distances larger than $1.5$, while the third term,
in which appeared the role of extra dimension, is always negative in
all distances and indicates that the extra dimension has
strengthened the attraction role in $F_{\rho}$. Moreover, the last
term, which displays the second contribution of the magnetic field,
is negative for $\rho\geqslant 0.5$. Overall, one can see that
always in large distances, $l_{z}$ has a repulsive effect and $J$
has an attractive effect in $F_{\rho}$. In other words, $J$ can be
considered as a source of "extra gravity". The combination effects
of all terms of $F_{\rho}$ is shown in Fig.\ref{rho12} and
Fig.\ref{rho1} in the absence and presence of the magnetic field.
Fig.\ref{rho2} displays that in $\theta=\dfrac{\pi}{2}$ the
attractive components of force are dominant in all distances, but
for smaller $\theta$ the repulsive components are dominant in some
distances. Fig.\ref{rho12} and Fig.\ref{rho22} illustrate how the
absence of magnetic terms impact on $F_{\rho}$. At the horizon
($\rho=1$), the $F_{\rho}$ is equal to $-[(l_{z}-b)^{2}+1+J^{2}]$.
So, the magnitude of $F_{\rho}$ at the horizon depends on the $b$,
$l_{z}$ and $J$, but the result is always negative. Therefore, the
massive particle at the horizon feels an attractive force along
$\rho$, in both the presence and absence of the magnetic field
(Fig.\ref{rho12} and Fig.\ref{rho1}). In large distances from the
black string and in the absence of a magnetic field, $F_{\rho}$
tends to zero ($\rho\rightarrow\infty \Longrightarrow
F_{\rho}\rightarrow 0$) and the massive particle does not feel any
forces along $\rho$ (Fig.\ref{rho12}). However, in the presence of
the magnetic field and in large distances, $F_{\rho} \rightarrow
-b^{2}(2\rho-1)$ which is always attractive and only depends on the
magnitude of the magnetic field (Fig.\ref{rho1}). Overall, in the
defined situation, the interesting point is that increasing the
magnitude of $J$, decreases the radial interval in which the
$F_{\rho}$ is positive (repulsive radial force). For instance, for
$l_{z}=4$ and $J \geqslant 2$ there is no area in which $F_{\rho}
\geqslant 0$ ( Fig.\ref{rho12} and Fig.\ref{rho1}).

\begin{figure}[h]
\centering \subfigure[ ]{
\includegraphics[width=0.9\textwidth]{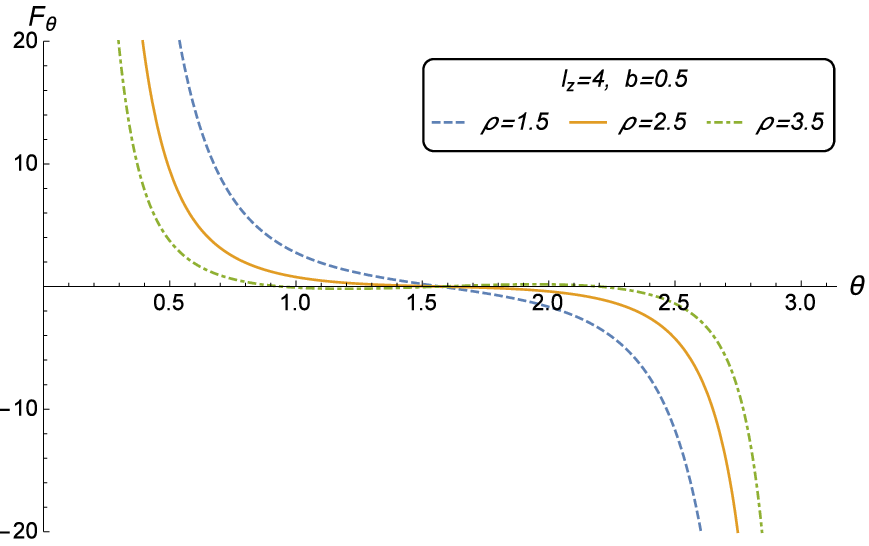}\label{theta1}
   }
\subfigure[ ]{
\includegraphics[width=0.9\textwidth]{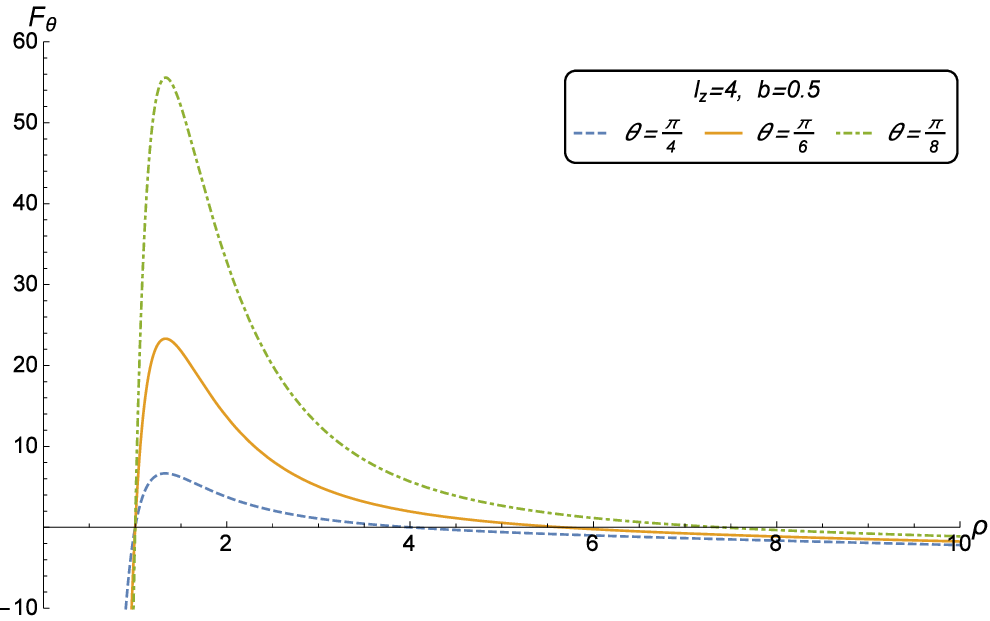}\label{theta2}
   }
\caption{(a) Effective Force $F_{\theta}$ versus $\theta$,
% in $l_{z}=4$ and $b=0.5$ for $\rho=1.5$ (blue dashed), $\rho=2.5$ (orange line),
%$\rho=3.5$ (green dotdashed),
(b) Effective Force $F_{\theta}$ versus $\rho$}
% in
%$J=2$ and $b=0.5$ for $\theta=\dfrac{\pi}{4}$ (blue dashed),
%$\theta=\dfrac{\pi}{6}$ (orange line), $\theta=\dfrac{\pi}{8}$ (green dotdashed).}
\label{ftheta}
\end{figure}

On the other hand, the polar component of effective force has
different features as follows: Firstly, as mentioned before, this
component ($F_{\theta}$) is independent of $J$. Secondly, since $b$
is supposed to be small and it appears in the second order, the
impact of a magnetic field on this component is really small and
negligible. As it is obvious from Eq.(\ref{(f-3)}) and
Fig.\ref{theta1}, for $\rho \geqslant 1$ and $ 0\leqslant \theta
\leqslant \dfrac{\pi}{2}$, $F_{\theta}$ is always repulsive (the
first term in Eq.(\ref{(f-3)}) is not large enough to affect), and
in the lower hemisphere ($ \dfrac{\pi}{2} \leqslant \theta \leqslant
\pi $) it is always attractive. The relation between $F_{\theta}$
and $\rho$ is also interesting. There is always a distance in the
range of $1 \leqslant \rho \leqslant 1.5$ in which the $F_{\theta}$
has its maximum while decreasing the $\theta$ increases the amount
of this maximum. By the way, the suitable situation for
investigation of the particle motion is in the equatorial plane
$(\theta= \frac{\pi}{2})$ in which this component of force is zero.
So, considering these features and since in $\theta=\dfrac{\pi}{2}$
the $F_{\theta}$ is zero, we chose $\theta=\dfrac{\pi}{2}$ plane to
investigate the escape velocity in the next section.

\section{Escape Velocity}
To find the escape velocity of a massive test particle which is
moving in the equatorial plane $(\theta_{0}= \frac{\pi}{2})$, the
collision between this particle and a particle that comes from
infinity is considered. After such a collision, the plane of the
first particle will tilt to $(\theta)$ plane. There are three
possible statuses, which depend on the mechanism of the collision
\cite{Zahrani:2013up}. When the transferred energy and momentum are
small, the orbit will perturb a little (bounded motion). However, if
the difference between the new and old total energy is large, the
particle can escape to infinity or fall into the black string. To
simplify the situation, following assumptions will be considered
\cite{Frolov:2010mi}, \cite{Zahrani:2013up}:

1. The initial radial velocity of the particle does not change
    $(\dot{\rho}_{0}=\acute{\dot{\rho}}=0)$.\\
2. The azimuthal angular momentum does not changed either,
\begin{equation}\label{(4-1)}
l_{z}=  l_{z0}= l,
\end{equation}
and the total angular momentum is \cite{Zahrani:2013up}:
\begin{equation}\label{(44-11)}
\textbf{l}^{2} \equiv  \rho^{4}
\dot{\theta}^{2}+\dfrac{l_{z}^{2}}{\sin ^{2}\theta }=\rho^{2}
v_{\bot}^{2}+\dfrac{l_{z}^{2}}{\sin ^{2}\theta }.
\end{equation}
Since the total angular momentum changes and it is assumed that
azimuthal angular momentum does not change, the particle gains a new
component of velocity which is orthogonal to the equatorial plane,
$v_{\bot}\equiv - \rho \dot{\theta}$ \cite{Zahrani:2013up}. If the
particle's energy after collision is larger than the effective
potential at infinity ($\varepsilon_{new}^2\geqslant 1+J^2$), the
escape condition will be satisfied and the particle will escape to
infinity ($v_{\bot}\rightarrow v_{esc}$).

Our additional assumption in five dimensions is:

3. The new constant of motion, the angular momentum related to the
fifth dimension is constant as well.

By using the first condition and putting the others in that, one can
find the escape velocities of neutral and charged particles.

\subsection{Neutral Particles}
Hence, for a neutral particle, by considering $b=0$ in
Eq.~(\ref{(2-4)}) and Eq.~(\ref{(2-5)}), we find:

\begin{equation}\label{(4-2)}
\varepsilon_{new}^2=\varepsilon_{0}^2+(1-\frac{1}{\rho_{0}})
v_{\perp}^2,
\end{equation}
in which:
\begin{equation}\label{(4-3)}
\varepsilon_{0}^2=(\frac{d\rho_{0}}{d\sigma})^2 + U_{eff},
\end{equation}
\begin{equation}\label{(4-4)}
U_{eff(\rho_{0})}=(1-\frac{1}{\rho_{0}})[\frac{l^2}{\rho_{0}^2}+
1+J^2],
\end{equation}
so:
\begin{equation}\label{(4-5)}
v_{\perp}=[\frac{\varepsilon_{new}^2}{(1-\frac{1}{\rho_{0}})}
-(\frac{l^2}{\rho_{0}^2}+1+J^2)]^\frac{1}{2},
\end{equation}
By imposing the escape condition:
\begin{equation}\label{(4-5-2)}
v_{esc}\geqslant[\frac{\varepsilon_{new}^2}{(1-\frac{1}{\rho_{0}})}
-(\frac{l^2}{\rho_{0}^2}+1+J^2)]^\frac{1}{2},
\end{equation}
in which:
\begin{equation}\label{(4-5-3)}
\varepsilon_{new}^2\geqslant 1+J^2,
\end{equation}
According to Eq.~(\ref{(2-16)}) and Eq.~(\ref{(2-17)}), the escape
velocity for a particle in ISCO is:
\begin{equation}\label{(4-5-2a)}
v_{esc(ISCO)}\geqslant\sqrt{\frac{1+J^2}{6}},
\end{equation}
which only depends on the $J$. In the case of $(w=0 \rightarrow
J=0)$, these results are similar to the results of
\cite{Frolov:2010mi}, \cite{Majeed:2014kka}, and
\cite{Zahrani:2013up} for Schwarzschild black hole.
In the asymptotic limit, %where $(\rho\rightarrow\infty)$%,
$\varepsilon_{new}^2\rightarrow 1+J^2 $ \cite{Majeed:2014kka}
obtains:
\begin{equation}\label{(4-6)}
v_{esc}=[\frac{1+J^2}{(1-\frac{1}{\rho_{0}})}-(\frac{l^2}{\rho_{0}^2}+1+J^2)]^\frac{1}{2}.
\end{equation}

\begin{figure}[h]
\centering \subfigure[ ]{
\includegraphics[width=0.9\textwidth]{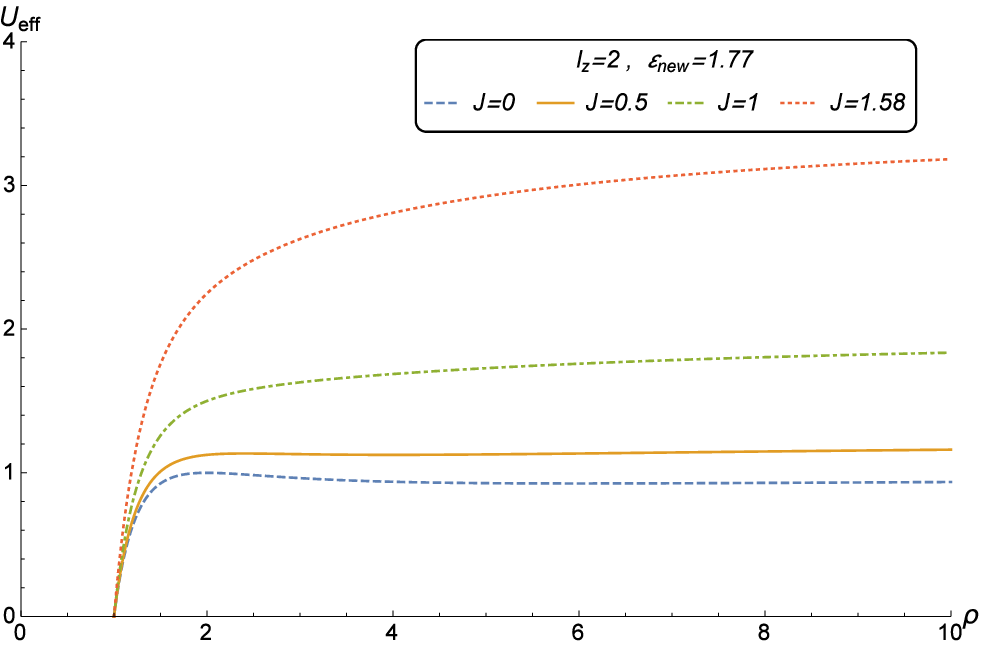}\label{a}
    }
\subfigure[ ]{
\includegraphics[width=0.9\textwidth]{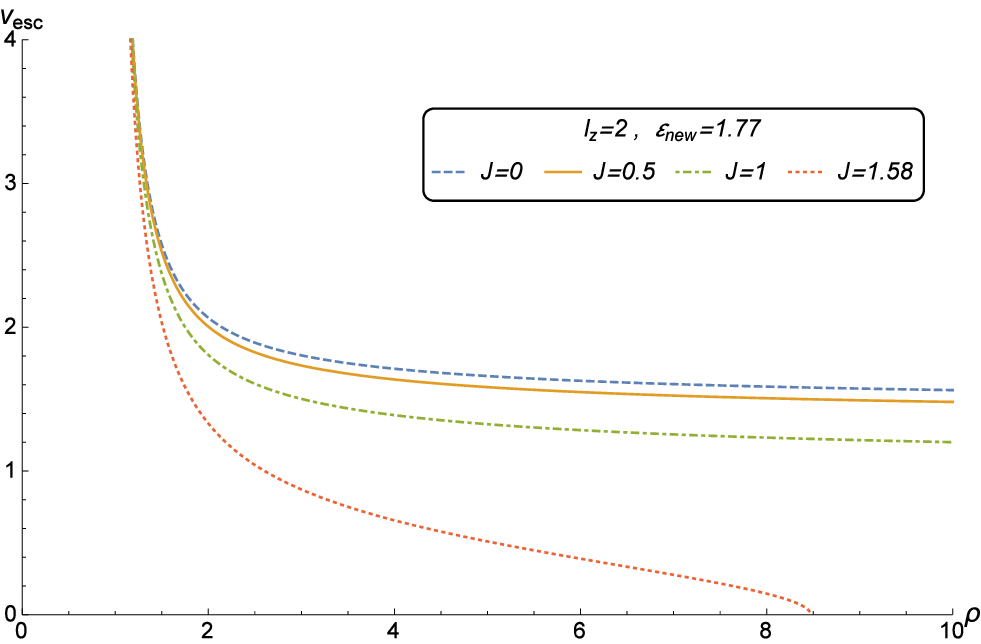}\label{b}
    }
\caption{Effective Potential and Escape Velocity versus $\rho$ for a
neutral particle and $l_{z}=2$, $\varepsilon_{new}=1.77$ in $J=0$
(blue dashed), $J=0.5$ (orange line), $J=1$ (green dotdashed),
$J=1.58$ (red dotted)} \label{f2}
\end{figure}

\begin{figure}[h]
\centering \subfigure[ ]{
\includegraphics[width=0.9\textwidth]{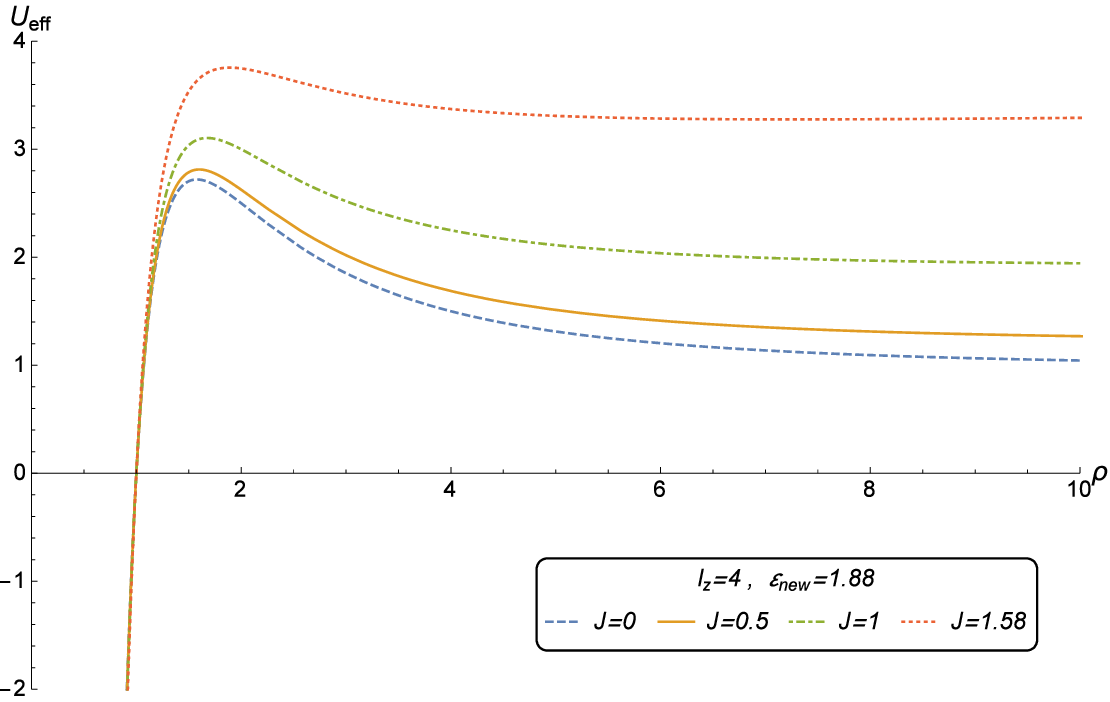}\label{c}
    }
\subfigure[ ]{
\includegraphics[width=0.9\textwidth]{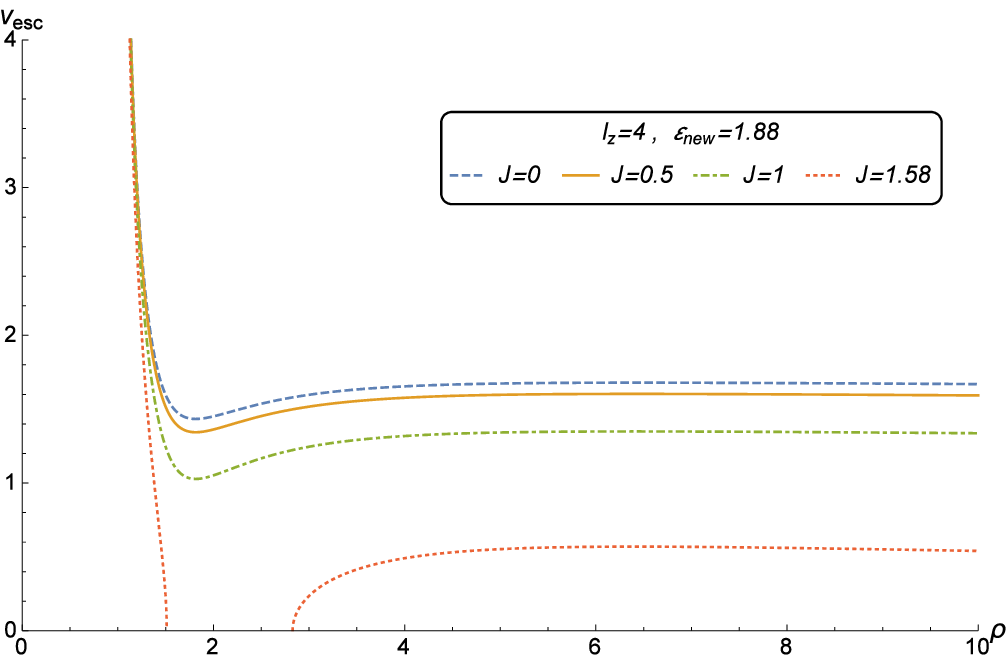}\label{d}
    }
\caption{Effective Potential and Escape Velocity versus $\rho$ for a
neutral particle and $l_{z}=4$, $\varepsilon_{new}=1.88$ in $J=0$
(blue dashed), $J=0.5$ (orange line), $J=1$ (green dotdashed),
$J=1.58$ (red dotted)} \label{f22020}
\end{figure}

\begin{figure}[h]
\centering \subfigure[ ]{
\includegraphics[width=0.9\textwidth]{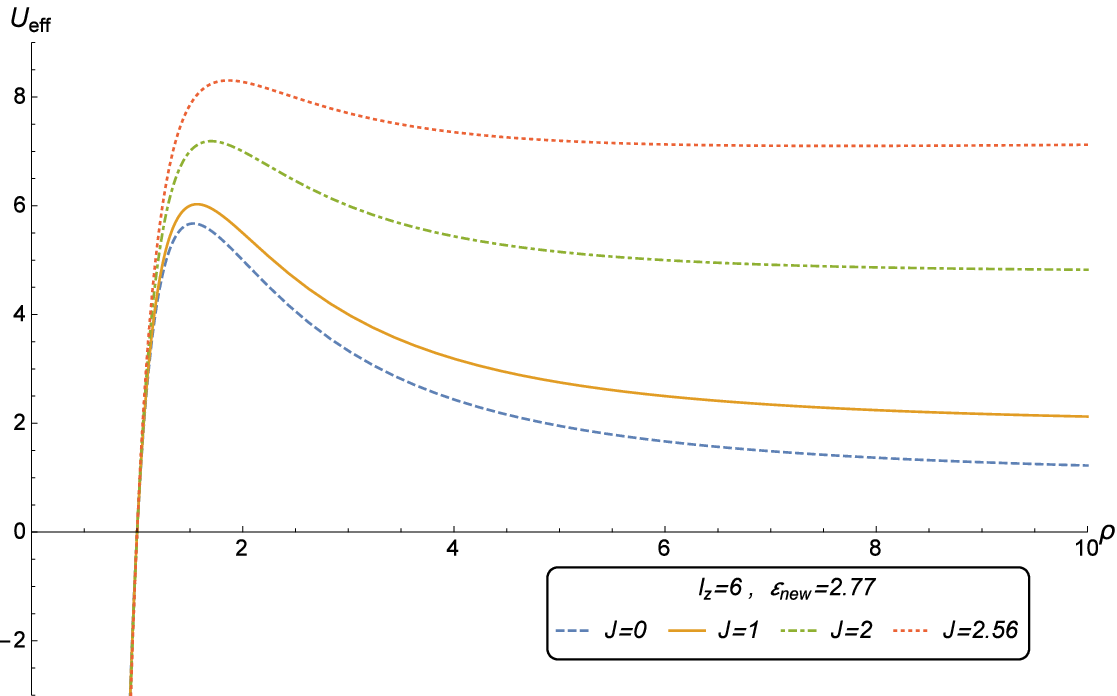}\label{e}
   }
\subfigure[ ]{
\includegraphics[width=0.9\textwidth]{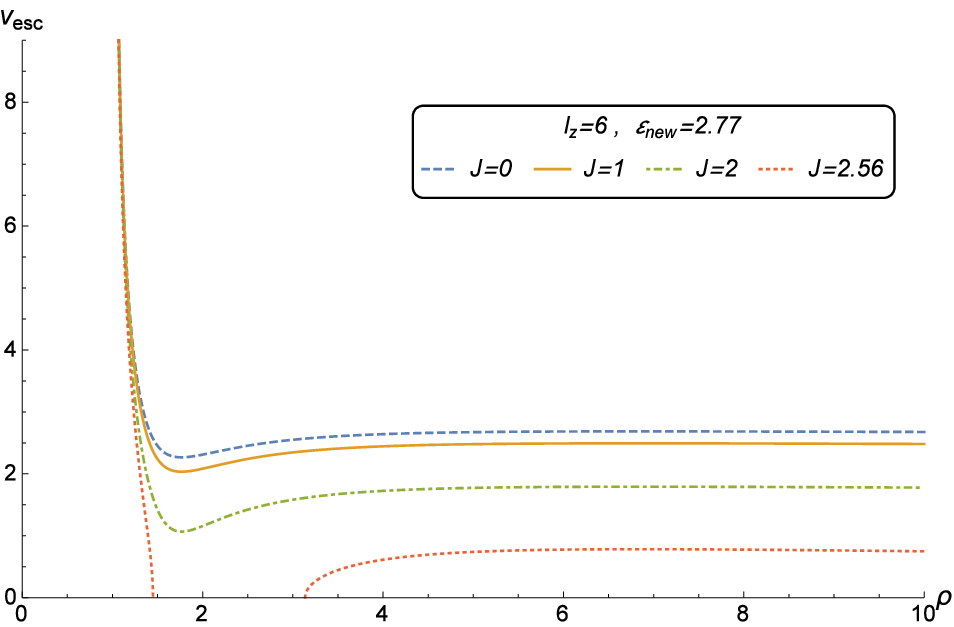}\label{f}
    }
\caption{Effective Potential and Escape Velocity versus $\rho$ for a
neutral particle and $l_{z}=6$, $\varepsilon_{new}=2.77$ in $J=0$
(blue dashed), $J=1$ (orange line), $J=2$ (green datdashed),
$J=2.56$ (red dotted).}
 \label{f220202}
\end{figure}

Fig.\ref{f2}, Fig.\ref{f22020}, and Fig.\ref{f220202} show the
relation between the escape velocity and the dimensionless distance
from the center of the black string and their relevant potential for
each case. According to Eq.~(\ref{(4-5-2)}) and as Fig.\ref{f22020}
and Fig.\ref{f220202} illustrate, the extra dimension shows its
effects in increasing the threshold of escape velocity in all
distances. Besides, in the large distances the escape velocity is
constant and comes from $\rho_{0}\rightarrow \infty$ in
Eq.~(\ref{(4-5-2)}). So, the constant magnitude of escape velocity
in the large distance is $(\varepsilon_{new}^2-1-J^2)^{1/2}$. Near
the event horizon ($\rho\rightarrow 1$), the escape velocity tends
to infinity ($v_{esc}\rightarrow+\infty$) which is expected
(Fig.\ref{b}, Fig.\ref{d} and Fig.\ref{f}).

Also, Fig.\ref{b} illustrates that for a small amount of $l$ and a
proper amount of $\varepsilon_{new}$, the escape velocity decreases
by the distance, which comes from the result of. Also, as it has
been shown in Fig.\ref{d} and Fig.\ref{f}, for large enough amounts
of $l$ and the proper amount of $\varepsilon_{new}$, there exists a
suitable $J$ (completely dependent on $\varepsilon_{new}$ by
$\varepsilon_{new}^2\approx 1+J^2$.) that vanishes the escape
velocity in some distance which means there is a balance between
attractive and repulsive forces. Moreover, as it is shown in
Fig.\ref{d} and Fig.\ref{f}, increasing the amount of $l$ causes
this special amount of $J$ to increase and also makes this distance
to be closer to the center of the black string. In addition, it is
obvious from Eq.~(\ref{(4-5-2)}) that in $J=0$ (Schwarzschild black
hole) such point does not exist in finite distances, as it has been
shown in Fig.\ref{d} and Fig.\ref{f}.

\subsection{Charged Particles}
\begin{figure}[h]
\centering \subfigure[ ]{
\includegraphics[width=0.9\textwidth]{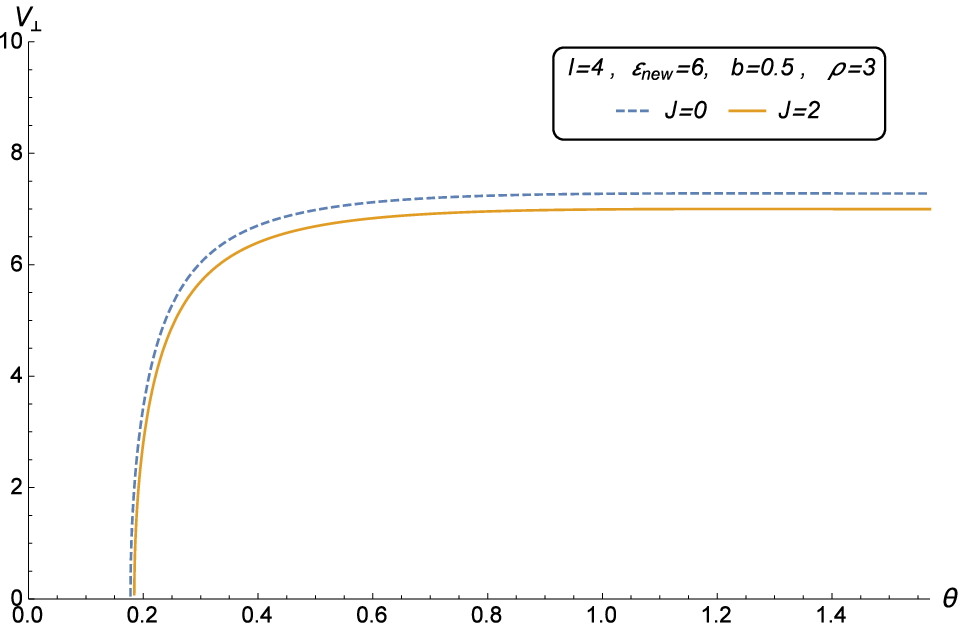}\label{fffa}
     }
\subfigure[ ]{
\includegraphics[width=0.9\textwidth]{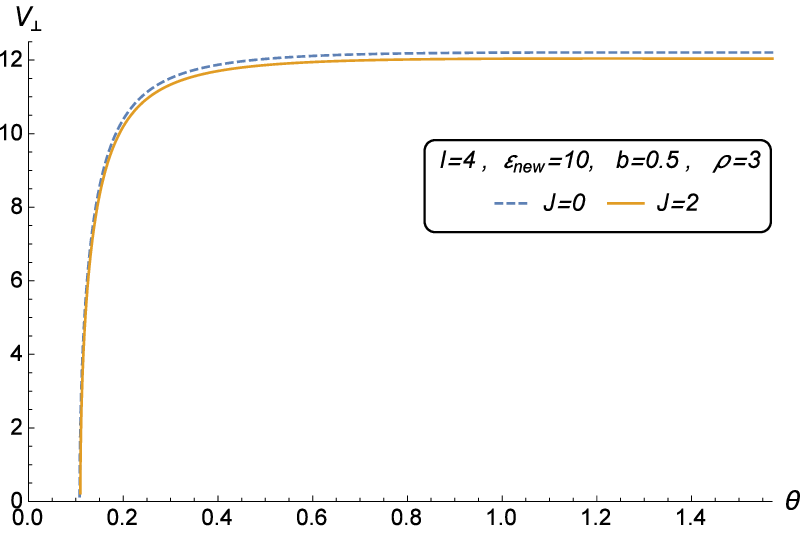}\label{c}
    }

\caption{$v_{\bot}$ versus $\theta$ for a charged particle in $l=4$,
$b=0.5$, $\rho=3$: (a) $\varepsilon_{new}=6$: $J=0$ (blue dashed),
$J=2$ (orange line), (b) $\varepsilon_{new}=10$: $J=0$ (blue
dashed), $J=2$ (orange line).} \label{fff}
\end{figure}

In the case of a charged particle, the situation is much more
complicated. In general by using all assumptions and by considering
Eq.~(\ref{(2-4)}) and Eq.~(\ref{(2-5)}), the result is:
\begin{equation}\label{(4-7)}
v_{\bot}^2=\dfrac{\varepsilon_{new}^2}{(1-\frac{1}{\rho_{0}})}
-1-J^{2}+2bl-\dfrac{l^2}{\rho_{0}^{2}\sin
^{2}\theta}-b^{2}\rho_{0}^{2}\sin ^{2}\theta,
\end{equation}
in which $\rho_{0}$ is the distance from the center of the black
string that according to the first assumption it does not change,
and the plane of particles tilts to $\theta$ after collision. So,
one can see that the result for a charged particle in a black string
is different from a black hole only in the existence of $J^{2}$ in
Eq.~(\ref{(4-7)}). Fig.\ref{fff} illustrates the impact of $J$ on
$v_{\bot}$. At the very first moment after collision $\theta$ is
$\dfrac{\pi}{2}$, so $v_{\bot}$ of a charged particle obtains from:
\begin{equation}\label{(4-72)}
v_{\bot}=\sqrt{\dfrac{\varepsilon_{new}^2}{(1-\frac{1}{\rho_{0}})}
-1-J^{2}+2bl-\dfrac{l^2}{\rho_{0}^{2}}-b^{2}\rho_{0}^{2}},
\end{equation}
which for $J=0$, is in accordance with \cite{Jamil:2014rsa}.

\section{Motion and trajectories of particles}
In this section, we will discuss the analytical solutions of the
equations of motion and plot some possible orbits for massive
particles in the black string space-time. We rewrite Eqs.
\ref{(1-4)} and \ref{(1-99)} using Eq.\ref{(2-1)} and the Mino time
($\rho^{2}d\gamma=d\sigma$)\cite{Mino:2003yg}, \cite{Grunau:2013oca}
as follows:
\begin{equation}\label{rqa}
(\dfrac{d\rho}{d\gamma})^{2}=(\varepsilon^2-J^2-1)\rho^4+(J^2+1)\rho^3-l^2
(\rho^2-\rho)  ,
\end{equation}
\begin{equation}\label{rqb}
(\dfrac{d\rho}{d\gamma})^{2}=-b^2\rho^6+b^2\rho^5
+(\varepsilon^2-J^2+2lb-1)\rho^4+(J^2-2lb+1)\rho^3-l^2 (\rho^2-\rho)
.
\end{equation}
\\

\begin{figure}[h]
\centering \subfigure[TO with parameters $ L=2$, $ E=2 $, $ b=0 $, $
J=0.5 $, $ a_{2}=-4 $, $ a_{3}=4 $, $ g_{2}=0.08 $, $  g_{3}=-2.87 $
.]{
\includegraphics[width=0.45\textwidth]{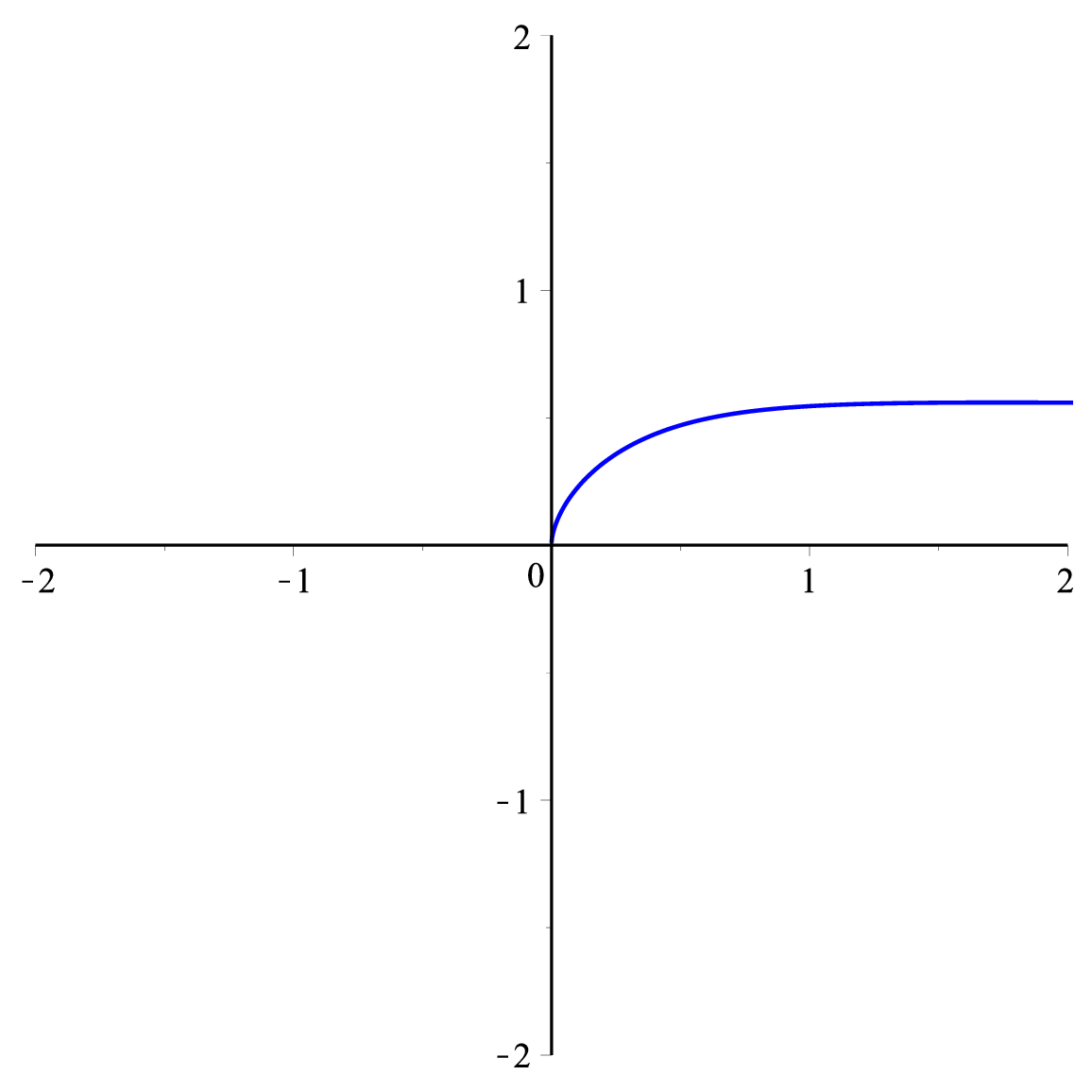}
    }
\subfigure[EO with parameters $ L=2.24$, $ E=1 $, $ b=0 $, $ J=0 $,
$ a_{2}=-5 $, $ a_{3}=5 $, $ g_{2}=0.84 $, $ g_{3}=0.06 $ .]{
\includegraphics[width=0.45\textwidth]{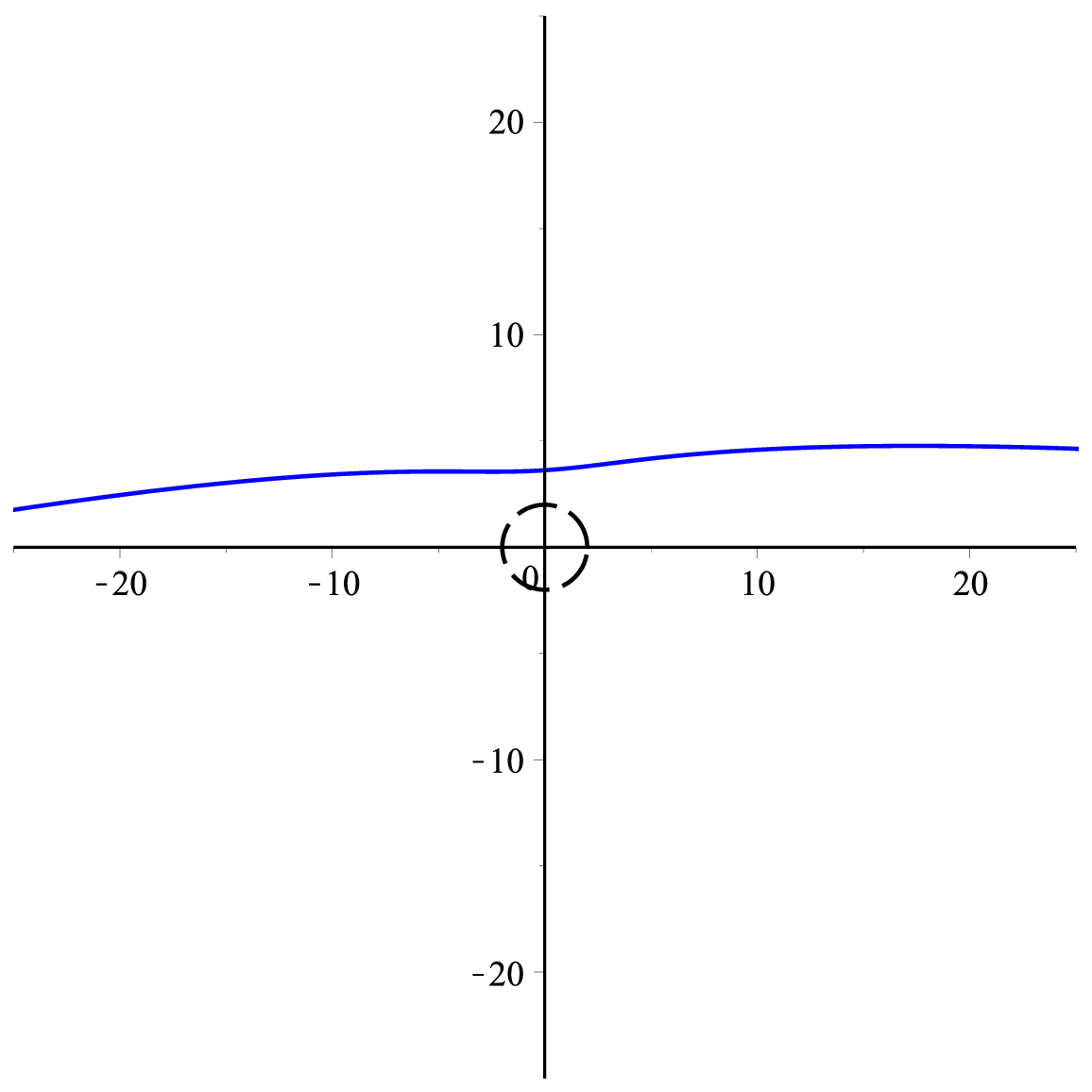}
    }
\subfigure[BO with parameters $ L=2.24$, $ E=1 $, $ b=0 $, $ J=0.2
$, $ a_{2}=-5 $, $ a_{3}=5 $, $ g_{2}=0.79 $, $ g_{3}=0.1 $.]{
\includegraphics[width=0.45\textwidth]{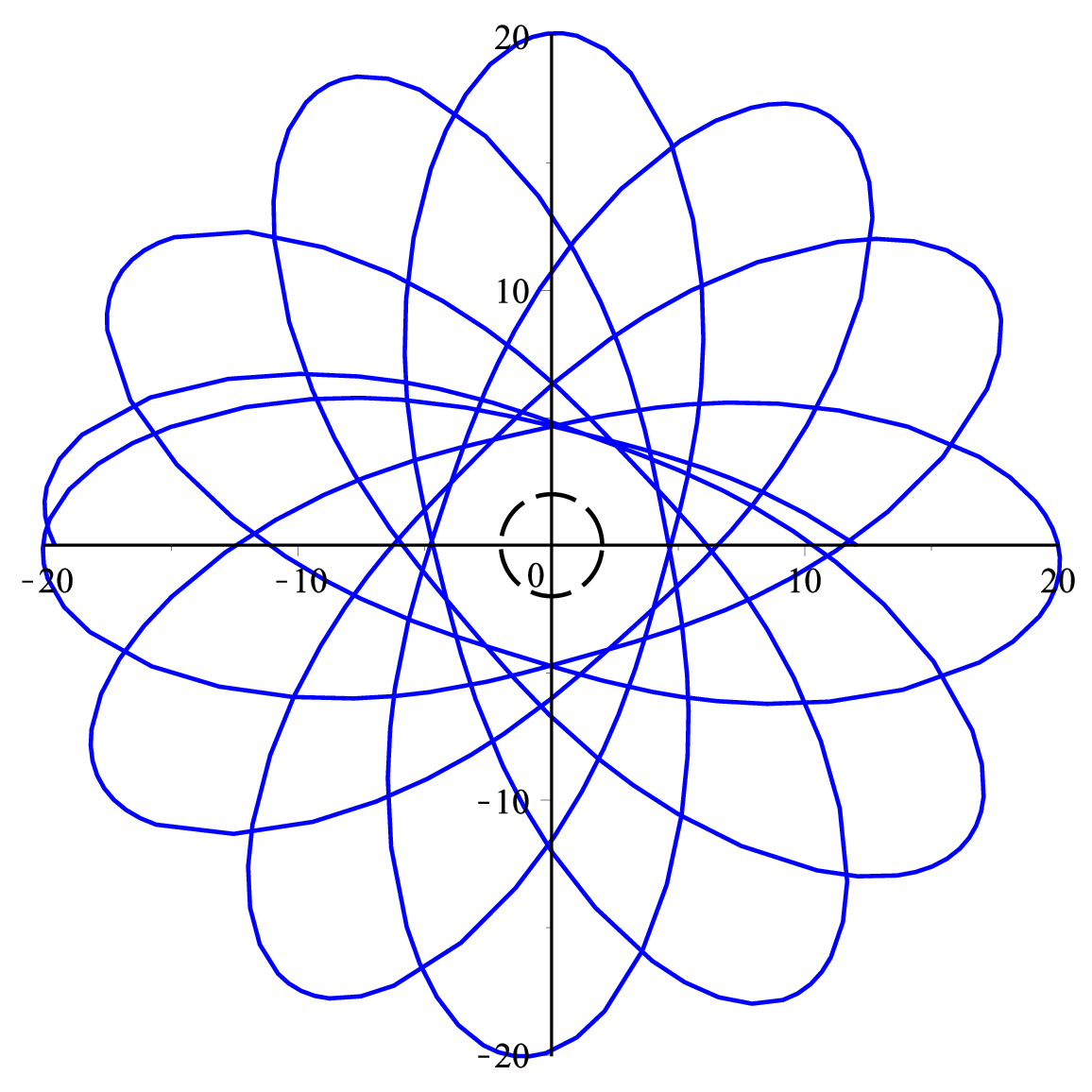}
    }
\caption{Three $xy$ plane plots of particle orbits in the black
string space-time in the absence of a magnetic field. The black
dashed circle depicts the position of the horizon and the blue
curves indicate the orbits.}\label{Orbit1}
\end{figure}

\begin{figure}[h]
\centering \subfigure[TO with parameters $ L=3$, $ E=1.25 $, $
b=0.01 $, $ J=0 $, $ a_{2}=-9 $, $ a_{3}=9 $, $ g_{2}=4.54 $, $
g_{3}=-1.22 $.]{
\includegraphics[width=0.34\textwidth]{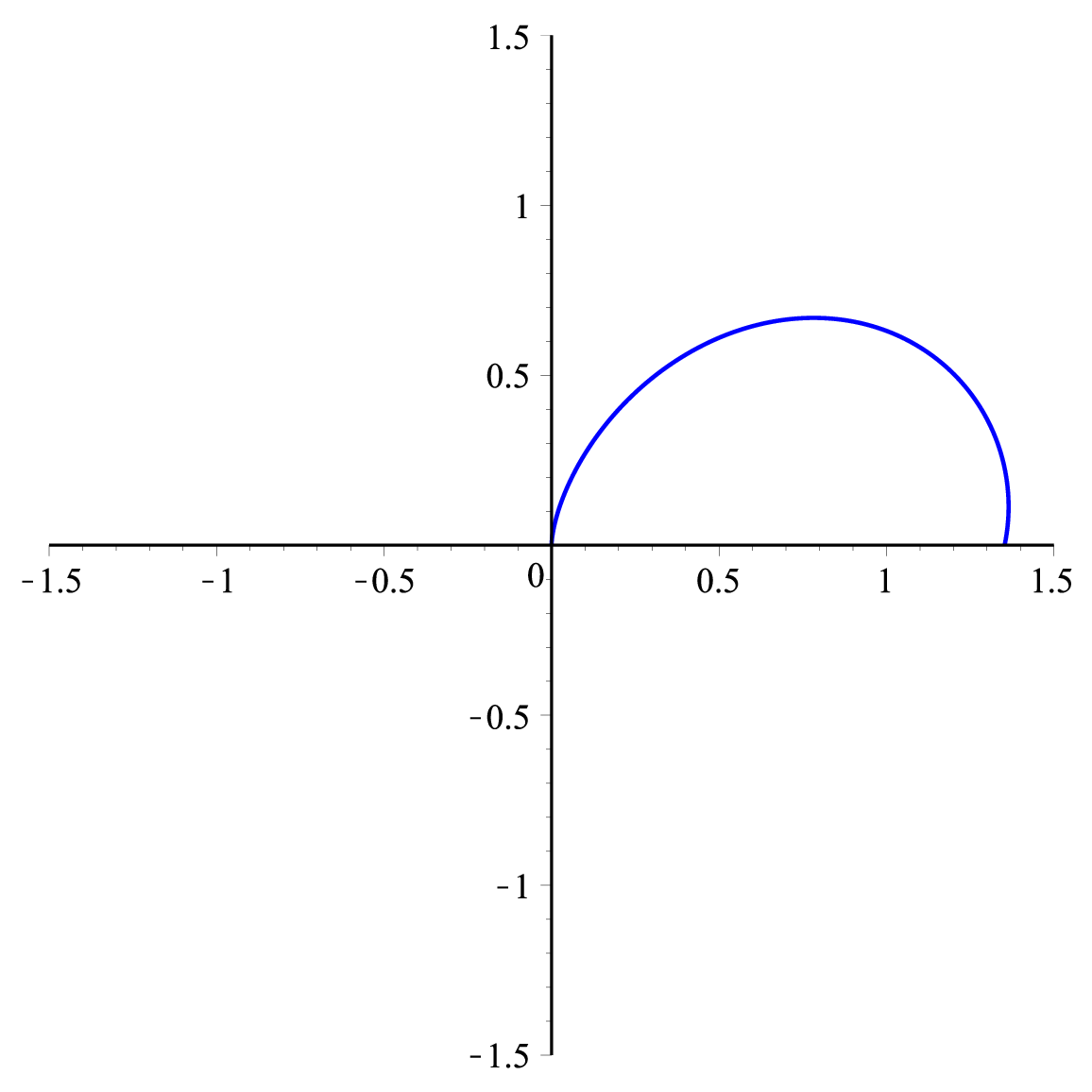}
    }
\subfigure[BO with parameters $ L=2$, $ E=0.97 $, $ b=0.01 $, $ J=0
$,  $ a_{2}=-4 $, $ a_{3}=4 $, $ g_{2}=0.37 $, $ g_{3}=-0.0046 $.]{
\includegraphics[width=0.34\textwidth]{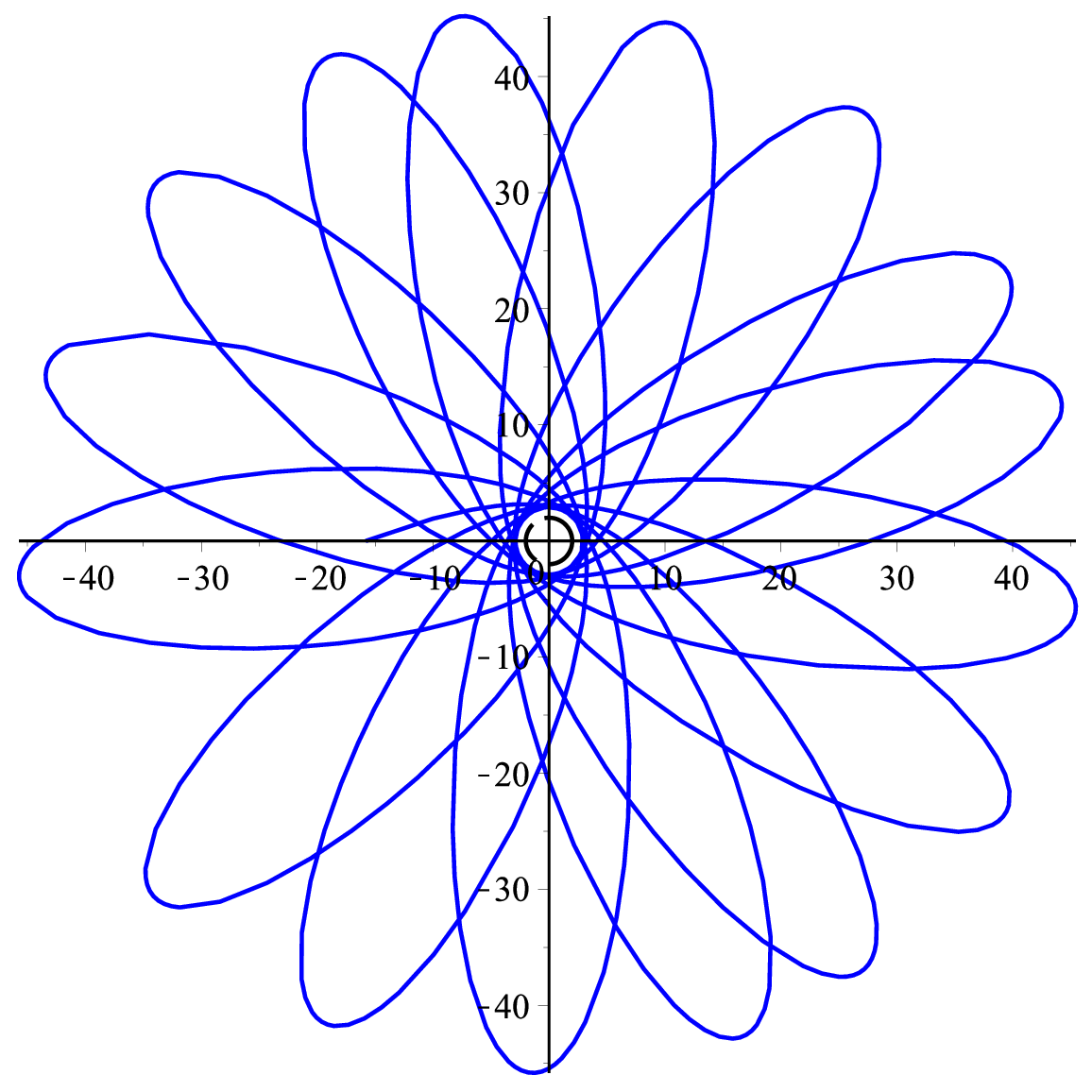}
    }
\subfigure[BO with parameters $ L=1.87$, $ E=0.97$, $ b=0.01 $, $
J=0.2 $, $ a_{2}=-3.5 $, $ a_{3}=3.5 $, $ g_{2}=0.14 $, $
g_{3}=-0.01 $.]{
\includegraphics[width=0.34\textwidth]{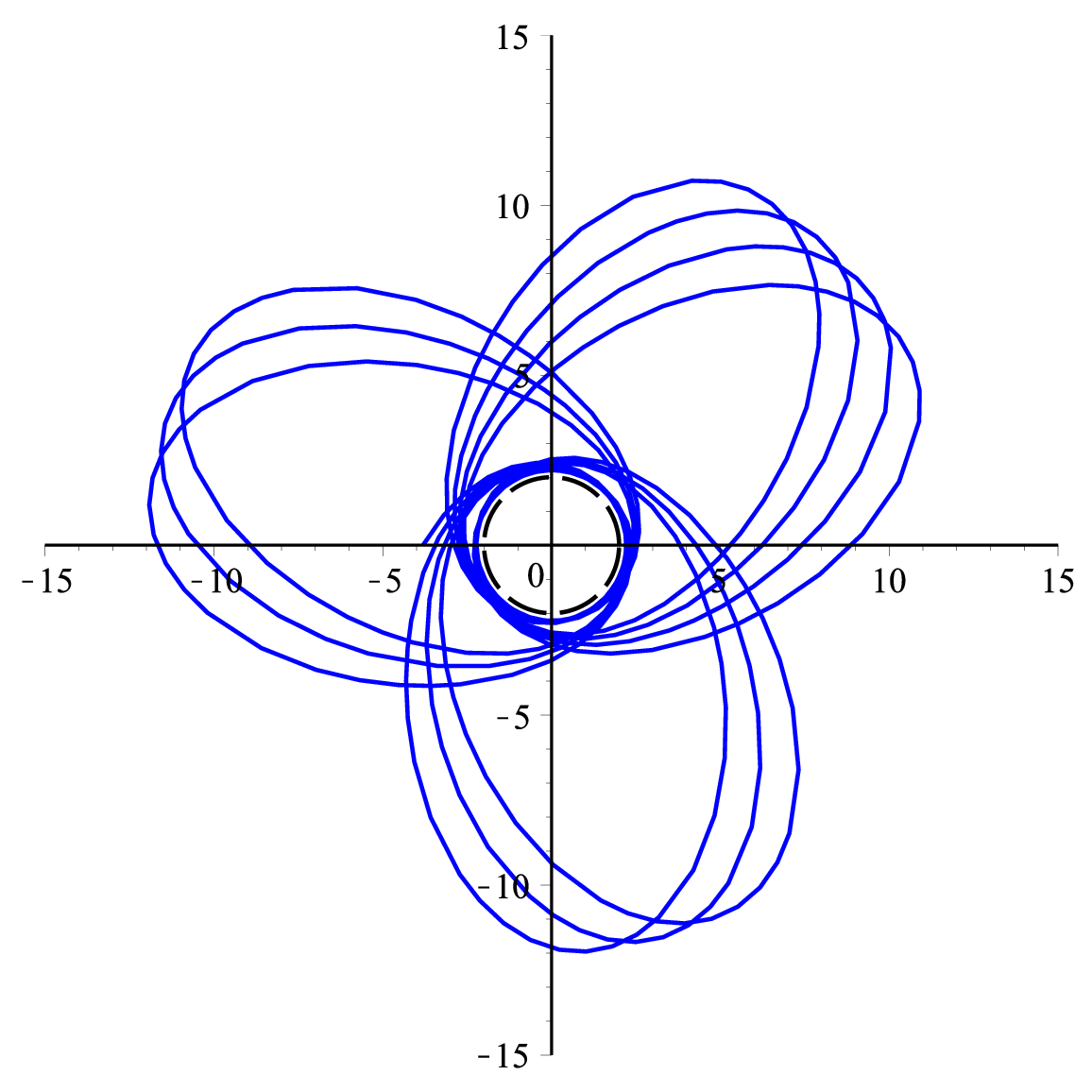}
    }
\subfigure[BO with parameters $ L=2$, $ E=0.97 $, $ b=0.01 $, $
J=0.2 $, $ a_{2}=-4 $, $ a_{3}=4 $, $ g_{2}=0.33 $, $ g_{3}=0.022
$.]{
\includegraphics[width=0.34\textwidth]{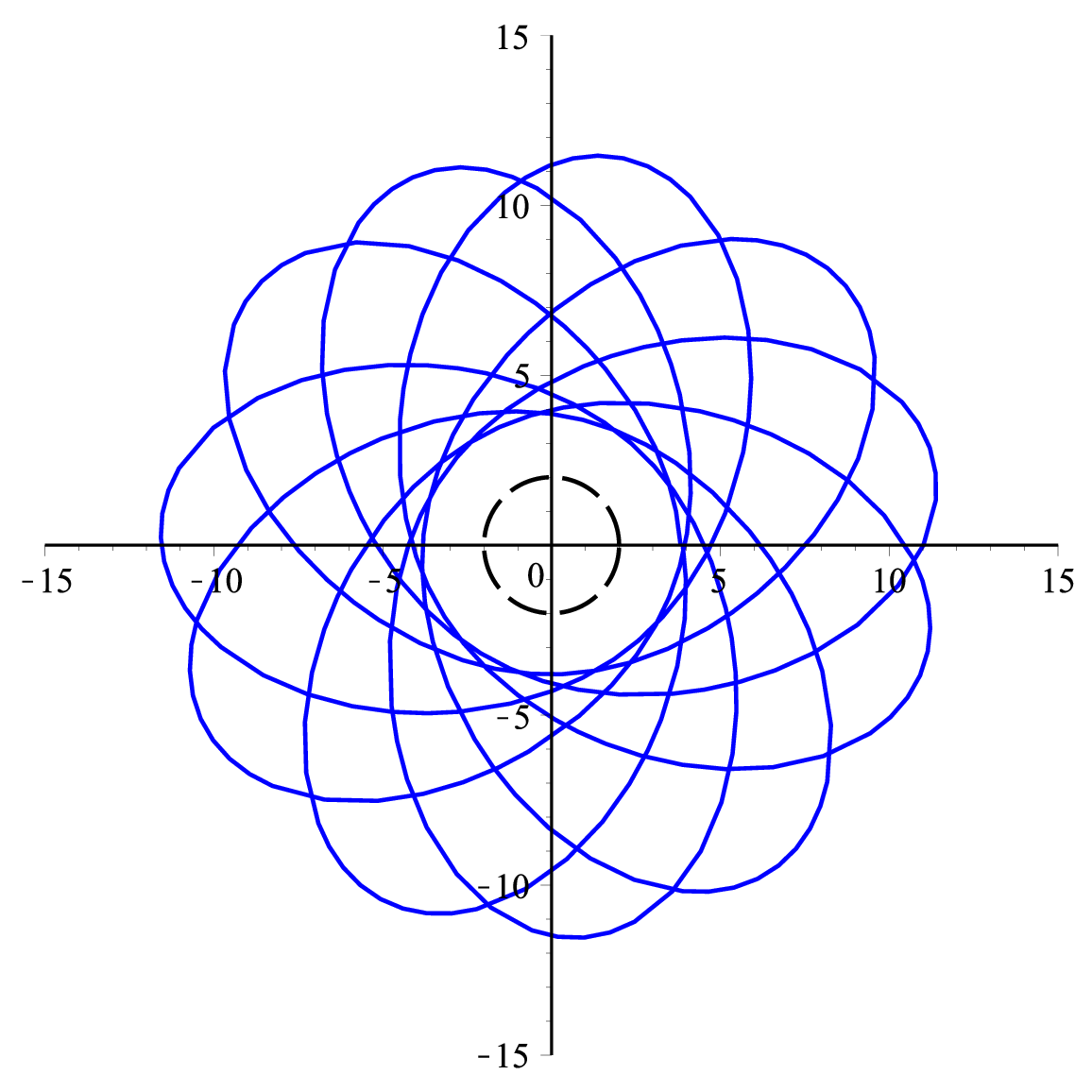}
    }
\subfigure[BO with parameters $ L=2$, $ E=0.97 $, $ b=0.006 $, $
J=0.2 $, $ a_{2}=-4 $, $ a_{3}=4 $, $ g_{2}=0.317 $, $ g_{3}=0.032
$.]{
\includegraphics[width=0.34\textwidth]{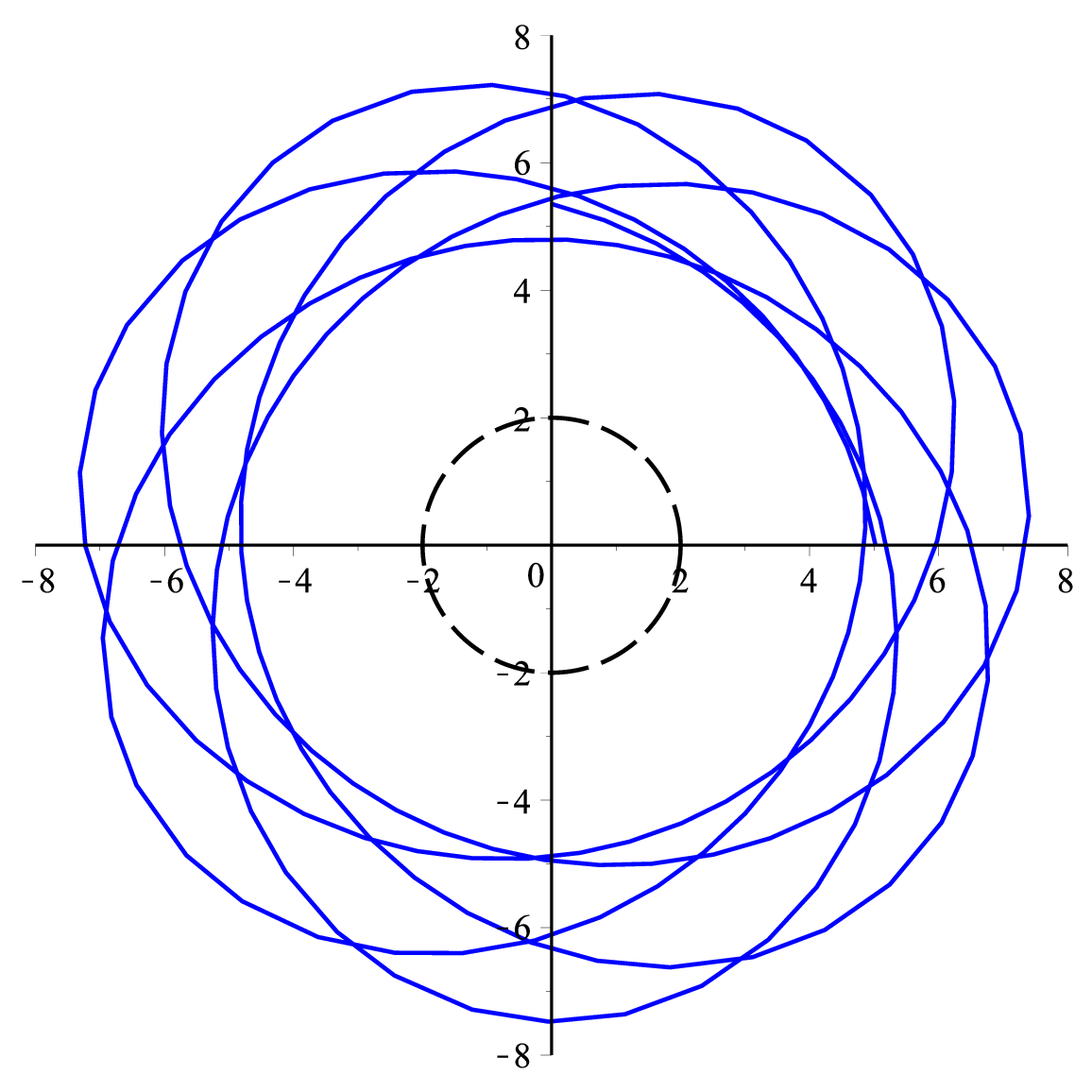}
    }
\caption{Five $xy$ plane plots of particle orbits in the black
string space-time in presence of a magnetic field. The black dashed
circle depict the position of the horizon and the blue curves
indicate the orbits.}\label{Orbit2}
\end{figure}

\paragraph{elliptic function}\label{seca}
Eq. (\ref{rqa}) and Eq. (\ref{rqb}) for $ b^{2}=0 $, are polynomials
of degree four as
$(\dfrac{d\rho}{d\gamma})^{2}=\sum_{i=0}^{4}c_{i}\rho^{i}$, and by
substitution $\rho=\frac{1}{u}+\rho_R$, can be convert to a
polynomial of degree three as:
\begin{equation}\label{deg3}
(\frac{du}{d\gamma})^2=\sum_{j=1}^3
a_ju^j,\;\;\;\;\;u(\gamma_0)=u_0,
\end{equation}
where $\rho_R $ is a zero of $R$ and:
\begin{equation}
a_j=\frac{1}{(4-j)!}\frac{d^{(4-j)}R}{d\tilde{r}^{4-j}}(\gamma_R),
\end{equation}
Again by substitution $u=\frac{1}{a_3}(4y-\frac{a_2}{3}) $,
Eq.(\ref{deg3}), transforms to an elliptical type differential
equation, known as the Weierstrass form as
\cite{Hackmann:2008zz,Soroushfar:2015wqa,Soroushfar:2016esy}:
\begin{equation}\label{wform}
(\frac{dy}{d\gamma})^2=4y^3-g_2y-g_3=p_3(y),
\end{equation}
with the Weierstrass invariants:
\begin{equation}
g_2=\frac{1}{16}(\frac{4}{3}a_2^2-4a_1a_3),\;\;\;
g_3=\frac{1}{16}(\frac{1}{3}a_1a_2a_3-\frac{2}{27}a_2^3a_3^2).
\end{equation}
Therefore, the solution of Eq. (\ref{wform}), using the Weierstrass
function, is:
\begin{equation}
y(\gamma)=\wp(\gamma-\gamma_{in};g_2,g_3),
\end{equation}
in which
$\gamma_{in}=\gamma_0+\int_{y_0}^{\infty}\frac{dy}{\sqrt{4y^3-g_2y-g_3}}$
with
$\gamma_0=\frac{1}{4}(\frac{a_3}{\rho_{0}-\rho_{R}}+\frac{a_2}{3})$
depends only on the initial value $\gamma_0$ and $\rho_{0}$. As a
result, the solution of polynomials of degree four (Eq. (\ref{rqa})
and Eq. (\ref{rqb}) for $ b^{2}=0 $) is
\begin{equation}\label{eqrw}
\rho(\gamma)=\frac{a_3}{4\wp(\gamma-\gamma_{in};g_2,g_3)-\frac{a_2}{3}}+\rho_R.
\end{equation}
\\
\paragraph{hyper-elliptic function}\label{secb}
Eq. (\ref{rqb}) is a polynomial of degree six as
$(\dfrac{d\rho}{d\gamma})^{2}=\sum_{i=0}^{6}c_{i}\rho^{i}$, which by
substitution $\rho=\frac{1}{y}+\rho_R$, transforms to a polynomial
of degree five as:
\begin{eqnarray}\label{twof}
\left(\frac{dy}{d\gamma} \right)
^2=P_{5}(y)=\sum_{i=0}^{5}a_{i}\rho^{i}.
\end{eqnarray}
The above equation is of a hyper-elliptic type, and so its analytic
solution can be written in the form of derivatives of the Kleinian
$\sigma$ function as \cite{Soroushfar:2016esy},
\cite{Hackmann:2008zz}, \cite{Enolski:2010if}:
\begin{equation}\label{deg5}
y(\gamma)=-\frac{\sigma_1(\gamma_{\infty})}{\sigma_2(\gamma_{\infty})}|_{\sigma(\gamma_{\infty})=0},
\end{equation}
where $ \gamma_{\infty}=(\gamma_2,\gamma-\gamma^{'}_{in}) $, $
\gamma^{'}_{in}=\gamma_{in}+\int_{\gamma_{in}}^{\infty}\frac{ydy^{'}}{\sqrt{P_5(y^{'})}}$,
and $\gamma_2$ is determined by the condition $
\sigma(\gamma_{\infty})=0 $. The other component in Eq.~(\ref{deg5})
is the function $\sigma_i$ which is the ith derivative of Kleinian
$\sigma$ function and $\sigma_z$ is:
\begin{equation}
\sigma_z=Ce^{zt}kz\theta[g,\theta](2w^{-1}z;T),
\end{equation}
in which  $T$ is the symmetric Riemann matrix, $C$ is a constant, $
\theta $ is the Riemann function with characteristic $[g, h]$ which
$2[g, h] =(0,1)^t+(1,1)^tT$, $\omega$ is the period matrix,
$k=\eta(2)^{-1}$ in which $\eta$ is the periodmatrix of the second
kind, therefore the solution of polynomials of degree six (Eq.
(\ref{rqb})) is:
\begin{eqnarray}\label{eqrsigma}
\rho(\gamma)=-\frac{\sigma_2(\gamma_{\infty})}{\sigma_1(\gamma_{\infty})}
.
\end{eqnarray}
\\
\paragraph{Trajectory}\label{orbit}
With these analytical results, we find some possible orbits for
particles in the black string space-time. Depending on $ b $, $J$,
$\varepsilon$ and $ l $ parameters, $ Terminating$ $orbit $ (TO), $
Escape $ $ orbit $ (EO) and $ Bound $ $ orbit $ (BO) are possible.
Examples of such particle orbits in the black string space-time in
the absence and in the presence of a magnetic field, are shown in
Figs.~\ref{Orbit1}, and \ref{Orbit2} respectively. It can be seen
from Fig.~\ref{Orbit1} that, in the absence of magnetic field, TO,
BO and EO are possible, where increasing the value of $J$, may
change an escape orbit to a bound orbit. While in the presence of
magnetic field, in which both BO and TO are possible (as depicted in
Fig.~\ref{Orbit2}) increasing the value of $J$, appears as
decreasing the radius of the bound orbit.

\section{Conclusion}
We studied some dynamical properties of neutral and charged
particles around a simple and weakly magnetized black string and the
attributes of their escape velocity. The results are:

i. Adding an extra dimension has a slight influence on the effective
potential, but the magnitude of the new constant of motion $(J)$
affects the shape of the potential and thereby the chance of
existence of the stable circular orbits.

ii. In the calculation of ISCO's parameters, up to the first order
of perturbation, due to the presence of a magnetic field, the radius
of ISCO is independent of the magnetic field and $J$, also the
solution resembles Schwarzschild black hole. But, $l_{zISCO}$ and
$\varepsilon_{ISCO}$ are related to $J$ even before the
perturbation.

iii. We also calculated the components of the effective force. The
new constant of motion related to the new dimension, adds a pure
attractive term on $F_{\rho}$. But it does not have any effects on
the other component of the effective force.

iv. Moreover, we calculated the escape velocity of a neutral
particle and saw how the magnitude of $J$ would vanish the escape
velocity in some $\rho$ for a neutral particle. It does not happen
in a Schwarzschild black hole ($J=0$).

v. Finally, we analysed the analytical solutions of the geodesic
equations in terms of elliptic and hyperelliptic functions. With
these analytical results, we have found some possible orbits for
particles in the black string space-time. These orbits illustrate
that, independent of the presence or the absence of a magnetic field
term, the effect of the fifth dimension, $J$, appears as a source of
extra gravity in a planetary motion. In the absence of a magnetic
field, TO, BO and EO are possible, while in the presence of a
magnetic field, BO and TO are possible.

In conclusion, by comparing a black hole to a black string, we could
see that the general form of the effective potential and equations
of motion are almost similar, however, in some properties related to
the escape velocity of a particle, the magnitude of the new constant
of motion causes some differences and even shows the presence of a
new term on the effective force.

\begin{acknowledgements}
We would like to thank the referee for his/her insightful comments.
\end{acknowledgements}

\clearpage

\clearpage

\clearpage

\clearpage

%\clearpage

%\clearpage

\clearpage

\bibliographystyle{amsplain}

\end{document}